\documentclass[%
 reprint,
nofootinbib,
 amsmath,amssymb,
 aps,
pra,
]{revtex4-1}
\usepackage{graphicx}% Include figure files
\usepackage{dcolumn}% Align table columns on decimal point
\usepackage{bm}% bold math 
\usepackage{color}
\usepackage[T1]{fontenc}
\usepackage{mathrsfs,amsfonts,dsfont}
\usepackage{amstext}
\usepackage{mathtools}
\usepackage{enumitem} 
\usepackage{rotating} % for diagram- descriptions
\usepackage{stmaryrd} % QECC brackets
 
\usepackage{braket}
\usepackage{bbm}
\input{Qcircuit.tex}		
\newcommand{\ZDZ}{\mathbb Z/D\mathbb Z}
\newcommand{\FD}{\mathbb F_D}
\newcommand{\e}{\mathrm{e}} 
\newcommand{\meterX}{\meter_{\hspace*{3mm}{_x}}}  % X measurement 
\newcommand{\Span}{\mathrm{span}}

\newcommand{\fig}{Fig.}
\newcommand{\eq}{Eq.}

\begin{document}
\preprint{APS/123-QED} 
\title{
Propagation of generalized Pauli errors in qudit Clifford circuits
}
\author{Daniel Miller}\email{daniel.miller@hhu.de}
\author{Timo Holz}
\author{Hermann Kampermann}
\author{Dagmar Bru\ss}
\affiliation{Institut f\"ur Theoretische  Physik III, Heinrich-Heine-Universit\"at D\"usseldorf, D-40225 D\"usseldorf, Germany}
\date{\today}

\begin{abstract} 
It is important for performance studies in quantum technologies to analyze quantum circuits in the presence of noise.
We introduce an error probability tensor, a tool to track generalized Pauli error statistics of qudits within quantum circuits composed of qudit Clifford gates.  
Our framework is compatible with qudit stabilizer quantum error-correcting codes.
We show how the error probability tensor can be applied in the most general case, and we demonstrate an error analysis of bipartite qudit repeaters with quantum error correction. 
We provide an exact analytical solution of the error statistics of the state distributed by such a repeater.
For a fixed number of degrees of freedom, we observe that higher-dimensional qudits can outperform qubits in terms of  distributed entanglement.

\end{abstract}
\maketitle
% ------------------------------------------------------------- ----------------- 
% -------------------------------------------------------------------------------- 
% -------------------------------------------------------------------------------- 
\section{\label{sec:1}\protect Introduction}  
 
%Introduction
%%%%%%%%%%%%%%%%%%%%%%%%%
%%%%%%%%%%%%%%%%%%%%%%%%%
%%% Motivation
Quantum computation and quantum communication are progressing fields with the prospect of faster computation~\cite{ ShorsAlgorithm97, GroversAlgorithm}  
and secure communication~\cite{BB84,EkertProtocol,SixState}
in comparison to their respective classical counterparts.  
Entangled quantum states are a key resource for quantum communication. 
The most promising approach to distribute entangled states among remote users are quantum repeaters~\cite{Briegel98, 3generations}. %[{\color {red}[van Loock?]}].
Potentially  fruitful candidates for units of quantum information are higher-dimensional quantum systems, so-called qudits, as they inherently possess multiple degrees of freedom while being implementable with single photons~\cite{Bruss02, Gisin02, Scarani10, Silberhorn15, Wong15, Gauthier17}.   

Often, quantum protocols are designed under the assumption of perfect control of the utilized  quantum systems.  Real experiments, however, are always subject to noise. 
This necessitates  studying such protocols in the presence of errors. 
% literature
In general, this problem is computationally hard since exponentially many classical resources are needed to simulate a quantum system. 
Explicit error analyses, however, have been carried out, e.g.,  for protocols based on qubits~\cite{BrownProbabilityVector, EppingAnalysis, EppingNetworks}.
In accordance to  the Gottesman-Knill theorem~\cite{GottesmanKnillTheorem}, 
this is possible due to the restriction to Clifford gates and Pauli error channels.
In~\cite{BrownProbabilityVector}, Janardan \emph{et al.} introduce a so-called error probability vector that allows to estimate the success probability of quantum protocols composed of Clifford operations in the presence of Pauli errors.

% This paper + notions in the title,
In this paper, we extend the applicability of this tool to qudits of fixed but arbitrary dimension $D\ge2$. 
For analytical investigations, it is helpful to rearrange its entries into a tensor, which we refer to as \emph{error probability tensor}.
To maintain compatibility with qudit stabilizer quantum error-correcting codes (QECCs),
we use the same  generalization of Pauli operators as in \cite{Gottesman1999, Ghreghoriu_Standard_form}. 
These generalized Pauli operators are unitary, traceless, and form an orthonormal basis for complex $D\times D$ matrices.  %~\cite[(2.65)]{WolfPhD}
Our error probability tensor provides a systematic procedure to track the statistics of generalized Pauli errors through quantum circuits composed of Clifford gates -- gates which transform generalized Pauli operators into one another.   
 
% outline
The paper is structured as follows.
In Sec.~\ref{sec:2}, we review the necessary background about qudits.
In Sec.~\ref{sec:3}, we define the error probability tensor and describe its use.
%we explain how to apply it to keep track of error statistics. 
In Sec.~\ref{sec:4}, we apply the error probability tensor for the error analysis of a qudit repeater line~\cite{EppingRouter}.
In Sec.~\ref{sec:5}, we conclude and give an outlook on future work.

\section{\label{sec:2} Setting}

In this section, the notation we will use throughout the paper is introduced. It covers basic quantum information processing with qudits.

\subsection{Physical and logical qudits}
A qudit is a quantum system with a Hilbert space of dimension $D\ge2$.  
Following \cite{QEC},  % [p. 379-381]
we label computational basis states  with elements in $\ZDZ=\{0,1,\ldots,D-1\}$, 
the ring  of integers modulo $D$. 
Qudit pure states are written as $z_0 \ket 0 + z_1 \ket 1 + \ldots + z_{D-1}\ket{D-1}$ with coefficients $z_j\in \mathbb C$,  $\sum_{j\in\ZDZ} \vert z_j\vert ^2 =1$.
Similarly, for pure $n$-qudit systems we have 
\begin{align}
  \ket{\psi} = \sum_{\mathbf j \in (\ZDZ)^n} z_{\mathbf{j}}\, \ket{\,\mathbf{j}\,},
\end{align} 
where the multi-qudit computational basis states $\ket {\mathbf j}$  are labeled by vectors $\mathbf{j}=(j_1, \ldots, j_n)$  in the free module  $(\ZDZ)^n$. 
In the special case where $D$ is a prime number, $\ZDZ$ is the same as $\mathbb F_D$, the finite field of order~$D$. 
If all qudits are measured in the computational basis, 
the measurement result is the vector $\mathbf j$ 
with probability $\vert z_\mathbf j \vert^2$.   
 
% QECC 
To correct errors, QECCs can be employed.
An $\llbracket n,k,d\rrbracket _D$ QECC encodes $n$ physical qudits into $k\le n$ logical qudits. The distance $d$ of the code is the minimal weight of an error that maps a codeword to a different codeword.
The number of single-qudit errors  which a QECC with distance $d$ can correct is $\lfloor (d-1)/2\rfloor$~\cite{QEC}. 
Stabilizer QECCs for higher-dimensional qudits, first introduced by Gottesman \cite{Gottesman1999}, have a logical code space stabilized by an abelian subgroup of the generalized Pauli group.  
In our error analysis we consider quantum polynomial codes
\cite{GottesmanSecret, AharonovBenOr08, KKKS06, Cross08} 
whose construction is outlined in Appendix \ref{app:QEC}.

\subsection{Quantum computation with qudits} 
Here we review the important classes of generalized Pauli gates and error channels, as well as qudit Clifford  gates  \cite{Gottesman1999}. 
Up to a global phase, the generalized Pauli operators on a single qudit are products of the unitary operators 
\begin{align}
 X := \sum_{k\in \ZDZ} \ket{k+1}\bra{k}   
\end{align}
 and
\begin{align}
  Z := \sum_{k\in \ZDZ} \omega^k \ket{k}\bra{k},
\end{align}
where $\omega:=\e^{2\pi i / D}$.
For $n$ qudits there are (up to a global phase) $D^{2n}$ different generalized Pauli operators each of which can be written as
\begin{align}\label{eq:pauli_standard_form}
X^\mathbf{r}Z^\mathbf{s}   
&:= \bigotimes_{i=1}^n X^{r_i}Z^{s_i} 
&=\sum_{\mathbf{k}\in (\ZDZ)^n} \omega^\mathbf{k\cdot s} \ket{\mathbf{k+r}}\bra{\mathbf{k}} 
\end{align}
for unique vectors $\mathbf{r,s} \in (\ZDZ)^n$, where $\mathbf{k\cdot s}= \sum_{i=1}^n k_is_i$ is the standard bilinear form, and $\mathbf{k+r}=(k_1+r_1, \ldots, k_n+r_n)$ is  the vector addition in $(\ZDZ)^n$.
Two generalized Pauli operators commute up to a phase,
\begin{align} \label{eq:pauli_commutation}
    (X^\mathbf{r}Z^\mathbf{s})(X^\mathbf{r'}Z^\mathbf{s'}) &= \omega^{ \mathbf{r'\cdot s-r\cdot s'}} (X^\mathbf{r'}Z^\mathbf{s'})(X^\mathbf{r}Z^\mathbf{s}).
\end{align} 
A generalized Pauli error channel $\mathcal F: \rho \mapsto \mathcal{F}(\rho)$ is a completely positive trace-preserving map with Kraus operators $\sqrt{f_\mathbf{r,s}}X^\mathbf{r}Z^\mathbf{s}$, 
\begin{align}\label{eq:f_coefficients}
\mathcal{F}(\rho) &= \sum_{\mathbf{r,s} \in (\ZDZ)^n} f_\mathbf{r,s} (X^\mathbf{r} Z^\mathbf{s}) \rho (X^\mathbf{r} Z^\mathbf{s} )^\dagger,
\end{align}
where $\sum_\mathbf{r,s}f_\mathbf{r,s}=1$.
This can be seen as the application of the Pauli operator $X^\mathbf{r}Z^\mathbf{s}$ to the state $\rho$ with probability $f_\mathbf{r,s}$. 
The $n$-qudit depolarizing channel (see Appendix \ref{app:depol}),
\begin{align}\label{def:depol}
\mathcal{F}_\text{dep}: \ \rho \longmapsto f \frac{\mathbbm 1}{D^n}+ (1-f) \rho,
\end{align}   
is such a generalized Pauli error channel with probabilities
\begin{align}  \label{eq:coefficients_dep}
f_\mathbf{r,s}= \begin{cases}  1-f + \frac{f}{D^{2n}} & \text{ if $\mathbf{r=s}=(0,\ldots,0)$}\\
\frac{f}{D^{2n}} & \text{ otherwise} \hspace{5em}.
\end{cases}
\end{align} 

The qudit Clifford group is the largest set of unitary operators which transform Pauli operators into one another, 
i.e., for every Clifford operator $U$  and all vectors $\mathbf{r,s}$ there are some vectors $\mathbf{r',s'}$ such that $U (X^\mathbf{r}Z^\mathbf{s})U^\dagger\propto X^\mathbf{r'}Z^\mathbf{s'}$ holds. 
An important single-qudit Clifford gate is the  {Fourier gate},
\begin{align}
 F:= \frac{1}{\sqrt D}\sum_{j,k\in \ZDZ} \omega^{jk}\ket{j}\bra{k},
\end{align}
which satisfies $FXF^\dagger = Z$ and $FZF^\dagger =X^{-1}$. 
For $D=2$, the Fourier gate equals the Hadamard gate $H={(X+Z)/\sqrt 2}$.
Another common single-qudit Clifford gate is the  {multiplication-with-$l$-gate},
\begin{align}
 M(l) := \sum_{k\in\ZDZ} \ket{kl} \bra{k},
\end{align}
where $l\in \ZDZ $ must be invertible such that $M(l)$ is unitary.
The controlled-$X$ and -$Z$-gates,
\begin{align}
       \mathrm{C}X :&= \sum_{k\in \ZDZ} \ket{k}\bra{k} \otimes X^k   
\end{align}
and 
\begin{align}
  \mathrm{C}Z :&= \sum_{k\in \ZDZ} \ket{k}\bra{k} \otimes Z^k,
\end{align}
are examples of important two-qudit Clifford gates.

\section{\label{sec:3}Tracking of error statistics}
 
Errors can originate from the malfunction of quantum gates.  
One can model a noisy quantum circuit by a sequence of ideal quantum gates $U_i$, each of which is followed by an error channel $\mathcal{F}_{i}$, as depicted in Fig.\ref{fig:err_pushed_end} (a). 
\begin{figure}[h!]   
\vspace{3mm} \centering \begin{minipage}{\columnwidth} 
\Qcircuit @C=.7em @ R=.4em  @! { 
&\gate{U_1} & \measure{\mathcal F_1} &  \multigate{1}{U_2}  &  \multimeasure{1}{\mathcal F_2}& \qw & \qw &\qw \\
&\qw &\qw & \ghost{U_2} & \ghost{\mathcal F_2} &    \multigate{1}{U_3}  & \multimeasure{1}{ \mathcal F_3} &\qw  \\ 
&\qw & \qw & \qw & \qw & \ghost{U_3}  & \ghost{\mathcal F_3}  & \qw
\gategroup{1}{2}{1}{3}{.5em}{--} 
\gategroup{1}{4}{2}{5}{.5em}{--} 
\gategroup{2}{6}{3}{7}{.5em}{--} 
}  \end{minipage} \vspace{.5em} 

(a) A noisy quantum circuit.  \vspace{1em}
\begin{center}  $\Updownarrow$ \end{center}
\begin{minipage}{\columnwidth}  
\Qcircuit @C=.7em @ R=.4em  @! { 
&\gate{U_1} &   \multigate{1}{U_2}& \qw       & \multimeasure{2}{\mathcal E} &\qw \\
&\qw        &\ghost{U_2}          & \multigate{1}{U_3} &  \ghost{\mathcal E} &\qw  \\ 
&\qw        & \qw                 & \ghost{U_3}        &  \ghost{\mathcal E} & \qw 
} \end{minipage} \vspace{.5em} 

(b) An equivalent interpretation of the noisy circuit. 
\caption{A quantum circuit (a) with noise modeled by error channels $\mathcal F_i$, after ideal unitary gates $U_i$, is mathematically equivalent to an ideal quantum circuit (b) followed by some error channel~$\mathcal E$. }
\label{fig:err_pushed_end} 
\end{figure} 
All errors propagate to the end of the circuit giving rise to a single error  channel~$\mathcal{E}$, cf. Fig.\ref{fig:err_pushed_end} (b), which describes the error statistics of the circuit as a whole.
In general, it is difficult to derive $\mathcal{E}$ from the $\mathcal{F}_i$.

Here, we develop a mathematical framework to calculate  the final error channel  $\mathcal E$  in the case of qudit Clifford gates $U_i$ and generalized Pauli channels $\mathcal F_i$.
We start with the definition of the error probability tensor, and, in the subsequent subsections, we describe  how to employ it for error analyses.

\subsection{Definition of the error probability tensor } 
Throughout this paper, we can consider the case where the error statistics of the qudits' state are given by some generalized Pauli error channel $\mathcal E$, i.e., the qudits are in an erroneous state  
\begin{align}  \label{eq:PauliErrorChannelE}
\mathcal E(\rho) &=   \sum_{\mathbf{r,s}\in (\ZDZ)^n} p_\mathbf{r,s} (X^\mathbf{r}Z^\mathbf{s}) \rho (X^\mathbf{r}Z^\mathbf{s})^\dagger,
\end{align}
instead of the desired state $\rho$, where the coefficients ${p_\mathbf{r,s}\ge0}$ sum to 1.
Inspired by the notion of error probability vectors \cite{BrownProbabilityVector}, we regard these $D^{2n} $ coefficients as entries of a tensor
\begin{align}\label{def:tensor}
P:=(p_\mathbf{r,s}) _{ \mathbf{r,s}\in (\ZDZ)^n} ,
\end{align} 
which has $2n$ indices, 
$\mathbf{(r,s)}=((r_1,\ldots,r_n),(s_1,\ldots,s_n))$. 
We call $P$ the \emph{error probability tensor}.  
The error statistics of the state $\rho$ are uniquely determined by the entries of this tensor.

\subsection{Updating the error probability tensor }
\label{sec:3B}   
In our approach it suffices to know how $\mathcal{E}$, or equivalently $P$, changes after each instance in a quantum circuit. Starting with the identity channel $\mathcal{E}=\mathrm{id}$  at the beginning of the circuit, one can track how the error statistics transform step by step, until the end of the circuit.
In this section, we present rules of how the error probability tensor is updated at every single step. 
Across qudit Clifford gates, its entries are permuted.
At generalized Pauli error channels, the entries are updated via a tensor equation.

\subsubsection{Qudit Clifford gates; permutations} 
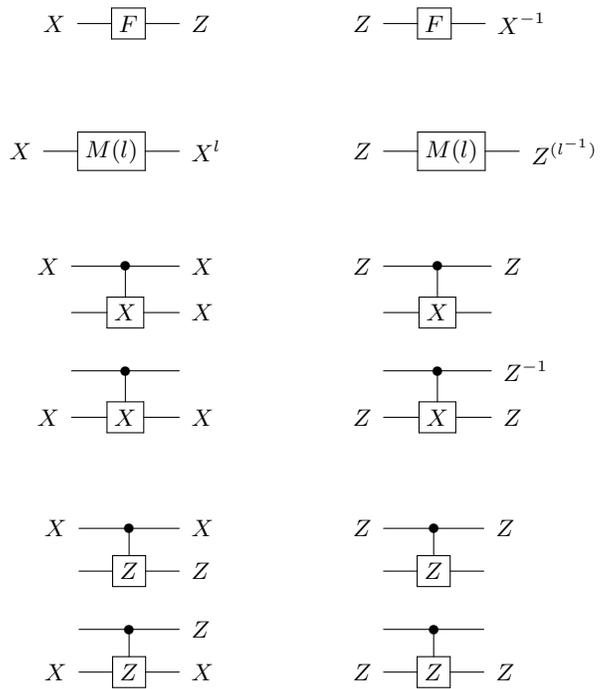
\begin{figure}[h!]
\begin{center} \begin{minipage}{.4\textwidth} \Qcircuit @C=.7em @ R=.4em  @!{ \lstick{X}  & \gate{F} & \rstick{ Z} \qw   } \end{minipage} \hspace{25mm}
\begin{minipage}{.4\textwidth} \Qcircuit @C=.7em @ R=.4em  @!{ \lstick{Z}  & \gate{F} & \rstick{X^{-1}} \qw   } \end{minipage}  \end{center}
\vspace{5mm}  
\begin{center}  \begin{minipage}{.4\textwidth} \Qcircuit @C=.01em @ R=.4em  @!{ \lstick{X}  & \gate{M(l)} & \rstick{X^{l}} \qw  } \end{minipage}
\hspace{25mm}
\begin{minipage}{.4\textwidth} \Qcircuit @C=.01em @ R=.4em  @!{ \lstick{Z}  & \gate{M(l)} & \rstick{Z^{(l^{-1})}} \qw   }
\end{minipage} 
\end{center}
\vspace{5mm}   
\begin{center}
\begin{minipage}{.4\textwidth} \Qcircuit @C=.7em @ R=.4em  @!{\lstick{X}  & \ctrl{1} & \rstick{ X} \qw \\   & \gate{X} & \rstick{ X} \qw  } \end{minipage} 
\hspace{25mm}
\begin{minipage}{.4\textwidth} \Qcircuit @C=.7em @ R=.4em  @!{\lstick{Z}  & \ctrl{1} & \rstick{ Z} \qw \\  & \gate{X} &   \qw } \end{minipage} \\
\vspace{5mm}
\begin{minipage}{.4\textwidth}  \Qcircuit @C=.7em @ R=.4em  @!{  & \ctrl{1} & \qw \\
\lstick{ X}   & \gate{X} & \rstick{  X} \qw  } \end{minipage}
\hspace{25mm}
\begin{minipage}{.4\textwidth} \Qcircuit @C=.7em @ R=.4em  @!{   & \ctrl{1} & \rstick{ Z^{-1}} \qw \\ \lstick{Z}   & \gate{X} & \rstick{Z} \qw  } \end{minipage}
\end{center} 
\vspace{5mm}
\begin{center} 
\begin{minipage}{.4\textwidth} \Qcircuit @C=.7em @ R=.4em  @!{ \lstick{X}  & \ctrl{1} & \rstick{X} \qw \\    & \gate{Z} & \rstick{Z} \qw  } \end{minipage} 
\hspace{25mm}
\begin{minipage}{.4\textwidth} \Qcircuit @C=.7em @ R=.4em  @!{ \lstick{Z}  & \ctrl{1} & \rstick{Z} \qw \\   & \gate{Z} &   \qw  } \end{minipage} \\
\vspace{5mm}
\begin{minipage}{.4\textwidth} \Qcircuit @C=.7em @ R=.4em  @!{  & \ctrl{1} & \rstick{Z} \qw \\ \lstick{X}   & \gate{Z} & \rstick{X} \qw } \end{minipage}
\hspace{25mm}
\begin{minipage}{.4\textwidth} \Qcircuit @C=.7em @ R=.4em  @!{  & \ctrl{1} &   \qw \\
\lstick{Z}   & \gate{Z} & \rstick{Z} \qw  } \end{minipage}
\end{center}  
\caption{Propagation rules of generalized Pauli errors for the $F$-, $M(l)$-, C$X$-, and C$Z$-gate. 
Across the $F$-gate, $X$ propagates into $Z$, and $Z$ into $X^{-1}$.
Across two-qudit gates, some single-qudit errors propagate into two-qudit errors, 
e.g., $X\otimes \mathbbm1 $ propagates into $X\otimes Z $ across the C$Z$-gate.
}
\label{fig:Clifford_propagation}   
\end{figure}

The propagation of single generalized Pauli errors across qudit Clifford gates is well known~\cite{Gottesman1999,Ghreghoriu_Standard_form}, cf. Fig.~\ref{fig:Clifford_propagation}. 
For the propagation of full error statistics, we use the fact that a qudit Clifford gate $U$
defines an automorphism  of $(\ZDZ)^{n}\times (\ZDZ)^{n}$, 
  $\pi_U: (\mathbf{r,s}) \mapsto (\mathbf{r',s'})$ via
\begin{align} 
 U (X^\mathbf{r} Z^\mathbf{s}) U^\dagger \propto  X^\mathbf{r'}Z^\mathbf{s'}.
\end{align} 
After the application of the (ideal) gate $U$, the error probability tensor $P=(p_\mathbf{r,s})$ is updated to 
\begin{align}
P'=(p'_\mathbf{r,s}) = (p_{\mathbf{r',s'}}).
\end{align} 
In other words, the entries of the error probability tensor are permuted.

\begin{figure}[h!]
\begin{minipage}{\columnwidth}  
 \centering
\hspace{-12mm}
\begin{minipage}{\columnwidth}  
\Qcircuit @C=.5em  @R=1em   {
\lstick{X^jZ^k} & &\qw & \ctrl{1} & \ctrl{1} & \qw& \qw & \ctrl{3}& \ctrl{3}   & \qw & \rstick{ X^j Z^{k+\mathbf{l\cdot b - m \cdot a}}} \\
 && \qw & \gate{X^{a_1}}& \gate{Z^{b_1}} & \qw & \qw &\qw &\qw &\qw  \\
\lstick{ X^\mathbf{l}Z^\mathbf{m}} &&  \vdots &&&&\ddots&&   &\vdots &\rstick{X^{\mathbf{l}+j\mathbf{a}}Z^{\mathbf{m}+j\mathbf{b}} }\\
 && \qw &\qw&\qw & \qw& \qw& \gate{X^{a_n}}& \gate{Z^{b_n}} &\qw 
 \gategroup{2}{2}{4}{2}{.7em}{\{}
 \gategroup{2}{3}{4}{10}{.7em}{\}}
}
\end{minipage} \vspace{3mm}
\caption{Across a sequence of C$X^{a_i}$- and C$Z^{b_i}$-gates,  the error
$X^j Z^k\otimes X^\mathbf{l} Z^\mathbf{m} $  propagates  
into $ X^j Z^{k+\mathbf{l\cdot b - m \cdot a}} \otimes X^{\mathbf{l}+j\mathbf{a}}Z^{\mathbf{m}+j\mathbf{b}} $. This defines the automorphism $\pi_{C(\mathbf{a,b})} $ in \eq\eqref{eq:sequence}. 
}
\label{fig:cpauli_propagation}   
\end{minipage} 
\end{figure}
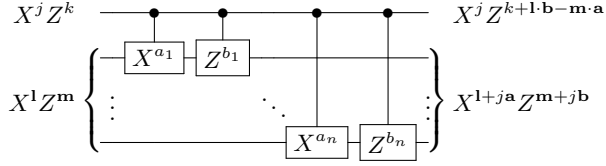

Now, we state explicit updating rules for the Clifford gates introduced in Sec.~\ref{sec:2}:
For every generalized Pauli gate $A=X^\mathbf{r}Z^\mathbf{s}$, the automorphism $\pi_A$ 
is the identity since Pauli operators commute up to a phase, recall \eq~\eqref{eq:pauli_commutation}. For other Clifford gates, $\pi_U$ might be nontrivial. 
For example for the Fourier gate $\pi_F(r,s)=(s,-r)$, and for the multiplication-with-$l$-gate $\pi_{M(l)} (r,s)=(l^{-1}r,ls)$.
Denoting by $C(\mathbf{a,b})$ the sequence of C$X^{a_i}$- and C$Z^{b_i}$-gates, cf. \fig~\ref{fig:cpauli_propagation},
we find
\begin{align}\label{eq:sequence}
\pi_{C(\mathbf{a,b})}&\left((j, \mathbf{l}), (k,\mathbf{m})  \right) \\
&= {((j,\, \mathbf{l}-j\mathbf{a}), ( k-\mathbf{l\cdot b+ m \cdot a},\, \mathbf{m}-j\mathbf{b}))}, \nonumber
\end{align}
where $j\mathbf{a}=(ja_1, \ldots, ja_n)$ is scalar multiplication in the module $(\ZDZ)^n$.

\subsubsection{Generalized Pauli channels; tensor equations}
\label{sec:3B2}
An $n$-qudit Pauli error channel $\mathcal F$ with coefficients $f_\mathbf{r,s}$, as   in \eq\eqref{eq:f_coefficients}, causes further errors. 
This is taken into account by updating the error probability tensor $P = (p_\mathbf{r,s})$ to $P'=(p'_\mathbf{r,s})$, where $p'_\mathrm{r,s}$ are the coefficients of the composed generalized Pauli error channel $\mathcal E' = \mathcal F \circ\mathcal E$,
\begin{align} 
\mathcal{E'}(\rho)&=   
\sum_{\substack{\mathbf{i,j,k,l,r,s} \in (\ZDZ)^n\\
\text{such that} \\ \mathbf{i+k=r}\text{ and } \mathbf{j+l=s} }} f_\mathbf{i,j}p_\mathbf{k,l}
(X^\mathbf{r}Z^\mathbf{s}) \rho (X^\mathbf{r}Z^\mathbf{s})^\dagger. \label{eq:fe}
\end{align}
Rewriting the sum and comparing to \eq~\eqref{eq:PauliErrorChannelE}, the entries of ${P'}$ are given by
\begin{align} \nonumber
p'_\mathbf{r,s} &= \sum_{\mathbf{k,l} \in (\ZDZ)^n}  f_\mathbf{r-k,s-l} \ p_\mathbf{k,l}  \\
&= \sum_{\mathbf{k,l}\in(\ZDZ)^n}  {F}_\mathbf{r}\,^\mathbf{k}\,_\mathbf{s}\,^\mathbf{l}  \ p_\mathbf{k,l},  
\label{eq:feKeinstein}
\end{align}
where $( F_\mathbf{r}\,^\mathbf{k}\,_\mathbf{s}\,^\mathbf{l} ) 
$ is a tensor 
with $2n$ covariant and $2n$ contravariant indices. Its entries are given by $ F_\mathbf{r}\,^\mathbf{k}\,_\mathbf{s}\,^\mathbf{l}:= f_\mathbf{r-k,s-l}$. 
This notation becomes very handy, when we deal with several error channels since we can abbreviate the last expression in \eq~\eqref{eq:feKeinstein} in the spirit of Einstein's sum convention as ${F}_\mathbf{r}\,^\mathbf{k}\,_\mathbf{s}\,^\mathbf{l}  \ p_\mathbf{k,l}$. 
%Such a product of tensors can be thought of as application of a stochastic matrix to a probability vector.

\subsection{Contractions of the error probability tensor}
In this section, we describe how contractions of the error probability tensor can be used 
to collect probabilities that correspond to similar events.  

\subsubsection{Measurements} 
If a qudit is measured in the computational basis, phase errors on that qudit become irrelevant.
Therefore, it is meaningful to add up the probabilities of all errors which only differ by $Z$-errors.
For example, suppose qudit $n$ is measured.
Then the error probability tensor is truncated to a tensor with $2n-1$ indices with entries 
\begin{align}
 p'_\mathbf{r,s'}  = \sum_{t\in \ZDZ} p_{\mathbf{r},(\mathbf{s'},t)},
\end{align}
where $\mathbf{s'}=  (s_1, \ldots, s_{n-1})$. 
After the contraction, the index $r_n$ is related to $D$itflip-errors on the measurement result, i.e., if $c$ was the correct outcome, the actual outcome is $c+r_n$ with conditional probability $p_\mathbf{r,s'}$ (conditioned on the presence of an $X^{(r_1,\ldots,r_{n-1})}Z^\mathbf{s'}$-error on the unmeasured qudits). 
This approach can be easily extended to the measurement of multiple qudits.
A measurement in the eigenbasis of a different Pauli operator can be substituted by an appropriate Clifford gate followed by a measurement in the computational basis.

\subsubsection{Discarding qudits} 
If, at some point in the analysis, one wants to keep track of errors on only $n'<n$ qudits (e.g., after discarding ancillas), one can trace out the error statistics of the $n-n'$ unnecessary qudits.
Assume w.l.o.g. that %qudits 1 to $n'$ are kept, and 
qudits $n'+1$ to $n$ are discarded.
The error statistics of the remaining qudits, stored in an
error probability tensor $P'=(p'_\mathbf{r',s'})$ with $2n'$ indices, %of type $(2m,0)$,
are given by
\begin{align} 
p' _\mathbf{r',s'} &= \sum_{r_{n'+1}, \ldots r_n, s_{n'+1}, \ldots s_n \in \ZDZ} p_\mathbf{r,s}, 
\end{align}
where $\mathbf r =(r_1,\ldots, r_n)$ is truncated to $\mathbf{r'}=(r_1,\ldots,r_{n'})$, and likewise for  $\mathbf{s}$ and $\mathbf{s}'$.

\subsubsection{Adding up probabilities of equivalent errors}
So far, we have shown how the error probability tensor describes the performance of a studied quantum circuit, independent of its input state. 
If, however, one is interested in the error statistics of a particular stabilizer state, it is reasonable to consider equivalence classes of errors: 
The stabilizer group of an $n$-qudit state $\ket{\psi}$ is generated by $n$ independent Pauli operators $S_i\propto X^{\mathbf{a}_i}Z^{\mathbf{b}_i}$ with $\mathbf{a}_i,\mathbf{b}_i \in (\ZDZ)^n$.
The exponents of all stabilizer operators form a submodule 
\begin{align} \label{eq:submodule}
 W:= \Span_{\ZDZ}\left\{ (\mathbf{a}_i,\mathbf{b}_i) \ \big\vert \ i\in\{1,\ldots,n\} \, \right\}
\end{align}
of $V:= (\ZDZ)^n \times (\ZDZ)^n $.
By definition, $X^\mathbf{a}Z^\mathbf{b} \ket{\psi}\propto  \ket{\psi}$ holds for every $\mathbf{(a,b)}\in W$. Likewise, for a given~coset
\begin{align}
\mathrm{co} (\mathbf{r,s}) &:= \left \{  (\mathbf{r+a,s+b}) \ \big\vert \ (\mathbf{a,b})\in W  \right\} ,
\end{align} 
i.e., an element
in the quotient module $V/W$, every pair of representatives $\mathbf{(j,k),(j',k') \in \mathrm{co} (r,s)}$ satisfies 
\begin{align}
 X^\mathbf{j}Z^\mathbf{k} \ket{\psi} \propto 
 X^\mathbf{j'}Z^\mathbf{k'} \ket{\psi},
\end{align}
since $ X^\mathbf{j}Z^\mathbf{k} $ and $ X^\mathbf{j'}Z^\mathbf{k'}$ are the same up to a stabilizer of $\ket{\psi}$ (and a global phase).  
It is meaningful to not distinguish between such errors as they lead to the same erroneous state. 
Hence, the error probability tensor $P=(p_{\mathbf{r,s}})$, as given in \eq~\eqref{def:tensor}, can be reduced to a tensor
$\bar P =(p_{\mathrm{co} (\mathbf{r,s})})$ with $D^n$ entries %(one for each $\mathrm{co} (\mathbf{r,s})\in V/W$)
\begin{align}
 p_{\mathrm{co}(\mathbf{ r,s})} = \sum_{\mathbf{(j,k)}\in\mathrm{co}(\mathbf{r,s})} p_\mathbf{j,k}.
\end{align}
We use cosets as indices because each of them corresponds to a whole class of errors with the same effect.

Consider, for example the maximally entangled state 
\begin{align}  \label{def:maximally_entangled_state}
    \ket{\Psi}%_\mathrm{A,B}
    &:=  %\text{ \color{red} ToDo: find the state stabilized  by $X_AZ_B, Z_AX_B$ } 
   \frac{1}{D} \sum_{j,k\in \ZDZ} \omega^{jk} \ket{j}%_\mathrm{A}
   \otimes \ket{k}%_\mathrm{B}
   ,
\end{align}
where again $\omega= \e^{2\pi i/D}$.
The stabilizers of $\ket{\Psi}$ are
$S_1= X\otimes Z = X^{(1,0)}Z^{(0,1)}$ and $S_2= Z\otimes X = X^{(0,1)}Z^{(1,0)}$. 
Hence, the submodule $W$ in Eq.~\eqref{eq:submodule} is spanned by $(\mathbf{a}_1,\mathbf{b}_1)=((1,0),(0,1))$ and $(\mathbf{a}_2,\mathbf{b}_2)=((0,1),(1,0))$, i.e., 
$ W=\{\left((\lambda,\mu),(\mu,\lambda)\right) \ \vert\ \lambda,\mu \in \ZDZ \}$. 
The elements in $V/W$ can be expressed as  
$ \mathrm{co} \left((0,r),(0,s)\right) = \left\{ ((\lambda,\mu+r),(\mu,\lambda+s)) \ \big\vert\ \lambda,\mu\in\ZDZ \right\}$,   
where $r,s\in\ZDZ$. The probability for an $X^rZ^s$-error on the second qudit -- or equivalently an $X^{-s}Z^{-r}$-error on the first qudit  
--  is given by 
 \begin{align}\label{eq:contract_stabilizers}
  p_{\mathrm{co} ((0,r),(0,s))} = \sum_{\lambda,\mu \in \ZDZ} p_{(\lambda,\mu+r),(\mu,\lambda+s)}.
 \end{align}

It is also possible to update the error probability tensor in its truncated form, where the stabilizers -- and hence $W$ and $V/W$ --  have to be updated after every Clifford gate.
This approach is recommended for numerical treatments as it gives an advantage in execution time and memory. 
For analyses carried out by hand, however, we recommend to first compute $P$ for the whole circuit, and to truncate to $\bar P$ afterwards, since calculating with quotient modules and cosets can be cumbersome.

\section{\label{sec:4}Application: qudit repeater line}
The purpose of quantum repeater networks is the distribution of entangled states among remote users. 
The approaches to overcome the presence of noise in quantum repeaters are categorized into three so-called generations~\cite{3generations}. 
Third generation quantum repeaters have, compared to generation one and two, the advantage of fast one-way communication~\cite{Lukin09, Fowler10}.  
There, qudits are encoded with a QECC which is used to correct loss and operational errors at the repeater stations. 
An $\llbracket n,k,d\rrbracket _D$ QECC is optimal if it saturates the quantum singleton bound $2d-2+k\le n$~\cite{QEC}.   % [(6.22)]
Prominent examples of such codes are $\llbracket 2d-1,1,d\rrbracket _D$ quantum polynomial codes \cite{GottesmanSecret, AharonovBenOr08, KKKS06, Cross08}, where $D$ is a prime and $d\le (D-1)/2$ is arbitrary. These are specified in Appendix~\ref{app:QEC}.

Using the error probability tensor, we carry out an error analysis of third generation quantum repeaters, cf. Appendices \ref{app:RL_ana}-\ref{app:RL_QEC_abortion}.
This is a generalization of the error analysis of qubit repeaters, performed in \cite{EppingAnalysis, EppingNetworks}, to the bipartite qudit case, which is a building block in qudit repeater networks \cite{EppingRouter}. 
In contrast to previous work \cite{EppingAnalysis, EppingNetworks, Abruzzo13, Loock17, Muralidharan17, Muralidharan18}, we do not compute secret key rates and certain cost functions.
Instead, we focus on deriving the full error statistics of the distributed state $\rho$, thus, $\rho$ itself.  Similar results are known for the qubit repetition code \cite{Bratzik14}.

In Sec.~\ref{sec:AliceNBob}, the ideal qudit repeater protocol is explained.
In Sec.~\ref{sec:RLQEC1}, we present an exact analytical solution of the error statistics of $\rho$ in terms of the qudit dimension $D$,  the number of repeater stations $N$, and   various error rates, $f_\mathrm{T}$ (transmission), $f_\mathrm{G}$ (C$Z$-gate), $f_\mathrm{M}$ (measurement)  and  $f_\mathrm{S}$ (storage). 
The distance $L_0$ between the repeater stations is only implicitly built-in via the transmission error rate $f_\mathrm{T}$. For example, optical fiber at telecommunication wavelengths has a channel loss of $0.2\mathrm{dB/km}$ \cite{Crypto02}.
In Sec.~\ref{sec:RLQEC2}, we discuss a quality-quantity trade-off of distributed states. 
Finally, in Sec.~\ref{sec:RLQEC3}, we compare the performance of $\llbracket 2d-1,1,d\rrbracket _D$ QECCs for variable code distance $d$ and physical qudit dimension $D$.

\subsection{The ideal qudit repeater line protocol} 
\label{sec:AliceNBob}   

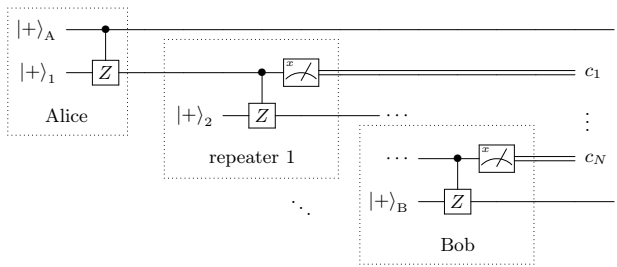
\begin{figure}[h!]
\begin{minipage}{\columnwidth}\centering 
\scalebox{0.8}
{ 
\Qcircuit @C=.2em @ R=.4em  @! {
&\lstick{\ket{+}_\mathrm{A}}&\ctrl{1}&\qw&\qw&\qw&\qw&\qw&\qw& \qw&\qw&\qw&\qw&\qw&\qw&\qw\\ 
&\lstick{\ket{+}_1}&\gate{Z}&\qw&\qw&\qw&\ctrl{1}& \meterX &\cw&\cw&\cw&\cw&\cw&\cw&\rstick{c_1}\cw\\
&\text{Alice}&&&&\lstick{\ket{+}_2}&\gate{Z}&\qw& \qw&\qw& \lstick{\cdots}&&&&\rstick{\vdots} \\
&&&&&&\hspace{-1em}\text{repeater 1}&&&& \lstick{\cdots}&\ctrl{1}&\meterX & \cw & \rstick{c_N} \cw \\
&&&&&&&\ddots&&&\lstick{\ket{+}_\mathrm{B}}&\gate{Z}&\qw&\qw&\qw&\qw \\
&&&&&&&&&&& \text{Bob}&&   &&&  & \			
\gategroup{1}{1}{3}{3}{2em}{.} % Alice 
\gategroup{2}{5}{4}{8}{2em}{.} % repeater1
\gategroup{4}{10}{6}{13}{2em}{.} % repeatern
}
} 
\end{minipage} 
\caption {A quantum circuit diagram representation of the qudit repeater line between Alice and Bob. 
Intermediate repeater stations are introduced to shorten the transmission distance of the qudits. All outcomes $c_i$ of the $X$-measurement at repeater $i$ are transmitted to Bob (who counts as repeater $N$) for the Pauli-frame recovery of $\ket \Psi $.
}
\label{fig:RL_perfect} 
\end{figure}

Consider two parties, Alice and Bob, both holding a single qudit. They want to create the maximally entangled state $\ket{\Psi}$ defined in Eq.~\eqref{def:maximally_entangled_state}.
To achieve this, Alice and Bob perform entanglement swapping via a {one-way} qudit repeater line. The protocol is as follows \cite{EppingRouter, Muralidharan17}.
Alice prepares two qudits (labeled $\mathrm{A}$ and $1$), each of them in the state $\ket{+}:=\frac{1}{\sqrt D}\sum_{j\in \ZDZ}\ket{j}$. She then applies a C$Z$-gate between these qudits which yields the state $\ket{\Psi}$ of \eq~\eqref{def:maximally_entangled_state}.  
Afterwards, she stores her qudit $\mathrm{A}$, and sends qudit $1$ to repeater station $1$.
There, another qudit (labeled $2$) is prepared in the $\ket{+}$ state. 
When qudit $1$ arrives, a C$Z$-gate is applied between qudits $1$ and~$2$.
Qudit 1 is then destructively measured in the $X$-basis.  
The measurement result is a classical digit $c_1 \in \ZDZ$.
Meanwhile, qudit $2$ is sent to the second repeater station, where the same steps as at station 1  are performed. 
Finally, after $N-1$ repeater stations, Bob receives qudit $N$, applies a C$Z$-gate to qudit $N$ and his own qudit (labeled~$\mathrm{B}$) and measures qudit $N$ in the $X$-basis. 
These steps are depicted in \fig~\ref{fig:RL_perfect}.

Alice and Bob now share a maximally entangled state whose exact form depends on all measurement outcomes $c_i\in\ZDZ$.  
Using the main-stabilizer approach of \cite{EppingRouter}, one can show that it is  the common $+1$ eigenstate of $\omega^{c_A}X_A \otimes Z_B$ and $\omega^{c_B}Z_A \otimes X_B$, where 
\begin{align}
c_\mathrm{A}:&= \sum_{i=1}^{N/2} (-1)^{i} c_{2i} 
\text{ \ and \ }
c_\mathrm{B} := \sum_{i=1}^{N/2} (-1)^{i} c_{N+1-2i}, 
\label{eq:postprocessed}
\end{align} 
and we assume that $N$ is even for simplicity. 
All classical digits $c_i$ are sent to Bob.
He post-processes them into $c_\mathrm{A}$ and $c_\mathrm{B}$, and applies the Pauli gate   $X^{c_\mathrm{A}}Z^{-c_\mathrm{B}}$ to his qudit $\mathrm{B}$.
Taking \eq~\eqref{eq:pauli_commutation} into account, this so-called Pauli-frame recovery produces the desired state $\ket{\Psi}$, as it is the unique two-qudit $+1$ eigenstate of $X_\mathrm{A}\otimes Z_\mathrm{B}$ and $Z_\mathrm{A}\otimes X_\mathrm{B}$.

\subsection{Error statistics of noisy qudit repeater lines} 
\label{sec:RLQEC1}
We now present analytical results for the error statistics of the third generation qudit repeater line described in the previous section. These results are valid for all polynomial codes and other $\llbracket n,1,d\rrbracket_D$ codes  with similar properties.  
The probability of a logical $X^rZ^s$-error on Bob's qudit $\mathrm{B}$, or equivalently 
of a logical $X^{-s}Z^{-r}$-error on Alice's qudit $\mathrm{A}$, recall Eq.~\eqref{eq:contract_stabilizers}, is 
\begin{align}
\label{eq:ana_sol}
 p_{\mathrm{co} ((0,r),(0,s))} &=    f^\mathrm{local}_{0,0}f^X_rf^Z_s + f^\mathrm{local}_\mathrm{err} \left( 1- f^X_rf^Z_s \right),
\end{align}
where ${f}^\mathrm{local}$ represent errors occurring locally on Alice's or Bob's qudit and $f^X$ and $f^Z$ represent errors propagating from repeater stations to the final state via Pauli-frame recovery.  
See Appendix~\ref{app:RL_ana} (without QECCs) and Appendix~\ref{app:RL_QEC} (with QECCs) for detailed derivations.

\begin{figure}[h!] \begin{center}
\begin{sideways} \hspace{4.4em} $p_{\mathrm{co} ((0,r),(0,s))}$ \end{sideways}
 \includegraphics[width= .9\columnwidth]{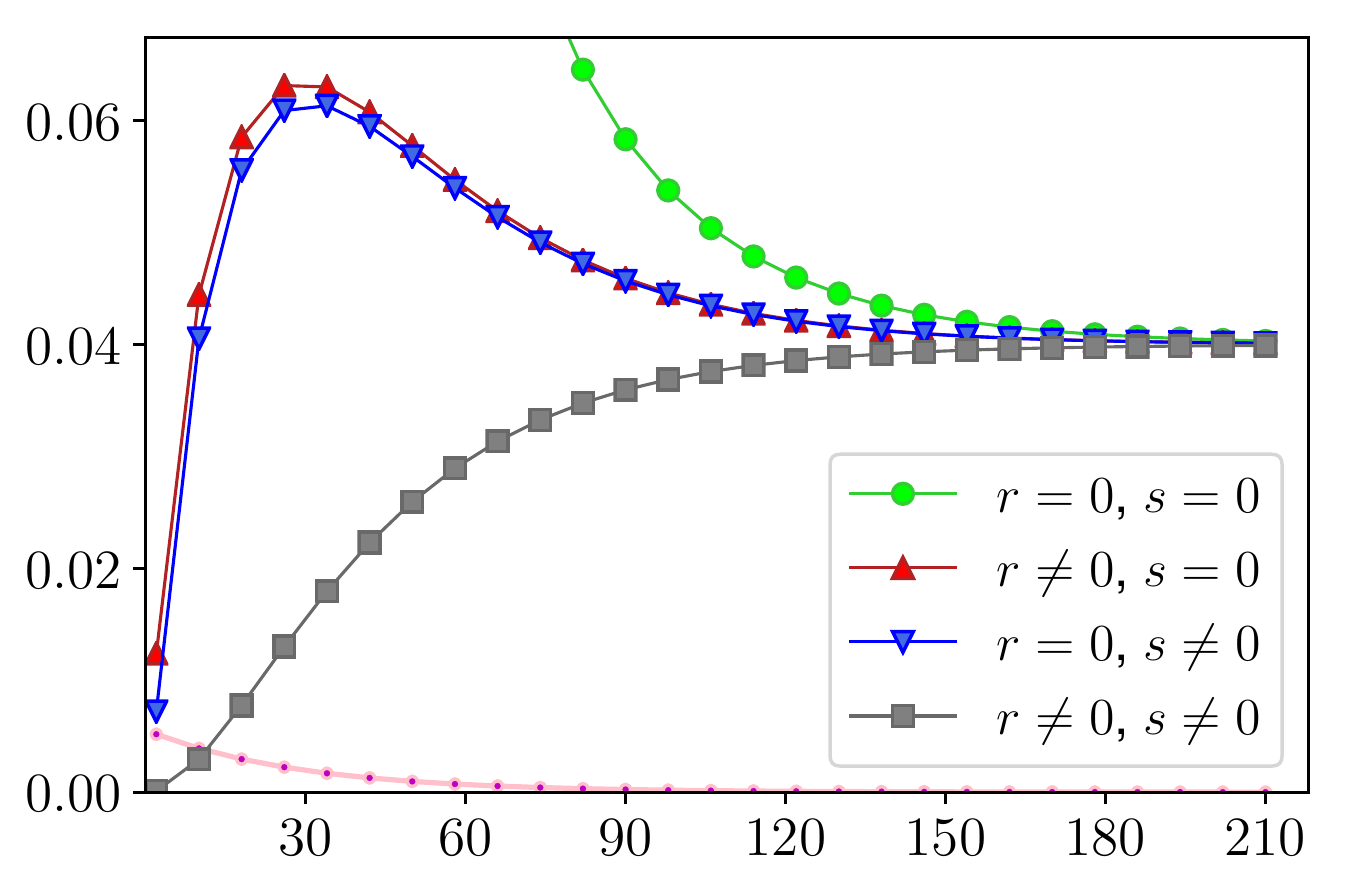}\\ 
\hspace{2.7em} $N$
\end{center}
\caption {Error statistics, \eq\eqref{eq:ana_sol}, of a state distributed with a ququint repeater line encoded with the $\llbracket 5,1,3\rrbracket_5$ quantum polynomial code.
The entries of the (reduced) error probability tensor are plotted as functions of the number of repeater stations, $N$,  
where transmission, measurement, gate, and storage  error rates are set to   $f_\mathrm{T}~=~0.05$, $f_\mathrm{M}=0.01$, $f_\mathrm{G}=0.001$,  and $f_\mathrm{S}=0.0001$, respectively. There are one green (circles; no errors), 4~red (triangles up; $X_\mathrm{B}^r$-error), 4~blue (triangles down; $Z_\mathrm{B}^s$-error), and 16 gray identical curves (squares; $X_\mathrm{B}^rZ_\mathrm{B}^s$-error). The pink curve (dots) shows the difference between red and blue curves.}
\label{fig:qec_tensor}   
\end{figure}

The error statistics for a 
fixed-error-rate\footnote{A transmission error rate of $f_\mathrm{T}=0.05$ corresponds to a repeater spacing of $L_0\approx 1\mathrm{km}$~\cite{Crypto02}. 
The best single photon detector efficiencies are about $95\%$~\cite{Hadfield09}, so we choose $f_\mathrm{M}=10^{-2}$ to keep the same order of magnitude.
Gate error rates are not known for qudit C$Z$-gates. 
We assume $f_\mathrm{G}=10^{-3}$ because this is the error rate of state-of-the-art single-qudit gates~\cite{Silberhorn15}, as well as a typical value for two-qubit gates in quantum communication~\cite{Briegel98, EppingAnalysis, EppingNetworks, Lukin09, Abruzzo13}.  
There are no good quantum memories yet. We still include storage errors with an optimistic assumption of $f_\mathrm{S}=10^{-4}$.}
ququint repeater line encoded with the $\llbracket 5,1,3\rrbracket _5$ quantum polynomial code  is plotted as a function of the number of repeater stations in \fig~\ref{fig:qec_tensor}. 
Note that we choose $D=5$ because it is the simplest case where a quantum polynomial code with code distance $d=3$ exists.

In total, there are $D^2$ curves in the figure, as curves of equal color overlap: 
One green, $D-1$ red, $D-1$ blue, and $(D-1)^2$ gray.
The green curve,
\begin{align}  
p_{\mathrm{co} ((0,0),(0,0))} &= \  1-\sum_{(r,s)\neq(0,0)}p_{\mathrm{co} ((0,r),(0,s))}, 
\end{align}
is uniquely determined by the other curves, and is directly related to the Uhlmann fidelity  $\sqrt{ \bra{\Psi}\rho\ket{\Psi} } = \sqrt{p_{\mathrm{co}((0,0),(0,0))} }$ of the distributed state $\rho$ \cite{Uhlmann76, Jozsa94}.
This curve decreases as a function of $N$ due to the fact that longer repeater lines contain more error sources, thus, mixing the state $\rho$. 
For $N\approx 200$ repeater stations,
$f^\mathrm{local}_{0,0}$ and $f^\mathrm{local}_\mathrm{err}$ converge to $1/D^2$, and $f^X_r$ and $f^Z_s$ converge to $1/D$, forcing all error probabilities to converge to an equilibrated value of $1/D^2=0.04$ as seen in Fig.~\ref{fig:qec_tensor}.
Hence, the distributed state approaches the maximally mixed state. 
The red and blue curves in the figure are the probabilities of single $X^r$- and $Z^s$-errors, respectively. 
These $D$it-flip and phase errors are introduced independently with probabilities $f^X_r$ and $f^Z_s$, respectively, which are  relatively small ($0< f^X_{r\neq0},f^Z_{s\neq0}\ll1/D$) for short repeater lines ($N<20$).
As a result, it is more likely that $D$it-flip and phase errors occur alone than together. This is why, for short repeater lines, the red and blue curves are much higher than the gray curves (simultaneous $X^rZ^s$-error). 
The red and blue curves surpass the equilibrium because accumulating $X$- and $Z$-errors have a low chance of canceling each other out.  
However, they can never surpass $1/D=0.2$ 
because the corresponding errors originate in $D$-outcome measurements at the repeater stations. 
There is an asymmetry in the probabilities of $X$- and $Z$- errors, as they accumulate differently at the ends of the repeater line. The difference between red and blue, which is plotted in pink, decreases in the repeater line's length, demonstrating the role of the finite-size of the repeater line. 
Note that, for qubits, the red and blue curves would not show a local maximum in $N$ since two $X$- and $Z$-errors, respectively, always cancel.

\subsection{Trade-off: fidelity vs. distribution probability}
\label{sec:RLQEC2} 

In practice, each qudit is encoded into the state of a photon.
During its transmission from one repeater station to the next, the photon is absorbed with probability
\begin{align}\label{def:f_abs}
 f_\mathrm{abs} &= 1- (1-f_\mathrm{C})\e ^{-\gamma},
\end{align}
where $f_ \mathrm{C}$ represents coupling losses, and the damping parameter $\gamma:= L_0/L_\mathrm{att}$ is the ratio of the repeater spacing $L_0$ to the attenuation length $L_\mathrm{att}\approx 20\mathrm{km}$ of the fiber through which the photon is transmitted~\cite{EppingAnalysis, EppingNetworks}. 
An error, which is caused by the absorption of the photon, is noticed by a non-click event at its measurement.
On the other hand, $f_\mathrm{T}$ represents unnoticed transmission errors. 
In the previous section, all errors were assumed to be undetected.
  
Similar to \cite{EppingAnalysis, EppingNetworks}, we consider a variation of the protocol. 
A measurement outcome is marked as ``?'' if an absorption of the corresponding photon is noticed. Such a lost qudit can be thought of as being in the completely mixed state, 
which is equivalent to  $X^rZ^s$-errors each with probability $1/D^2$. 
Hence, $Z^r$-errors are induced on the next qudit through the C$Z$-gate (recall \fig~\ref{fig:Clifford_propagation}), 
so the measurement of the next qudit has an error with probability $(D-1)/D$. 
As this is a high probability, we preventively also mark that measurement outcome as ``?''. 
The adapted strategy is to abort and restart the protocol if more than a fixed number $k_\mathrm{max}$ of measurement outcomes at a given repeater station have been marked as ``?''. 
If, however, only $k\le k_\mathrm{max}$ outcomes are marked as ``?'', they are discarded and the $n-k$ remaining outcomes form a classical error-correcting code with a Hamming distance of at least $d-k$. The logical measurement outcome is obtained by decoding the remaining physical outcomes according to this code.

\begin{figure}[h!]
\begin{center} 
\begin{minipage}{\columnwidth}
\begin{sideways} \hspace{5.5em} $F(k_\mathrm{max}) $ 
\end{sideways}
\includegraphics[width= .9\columnwidth]{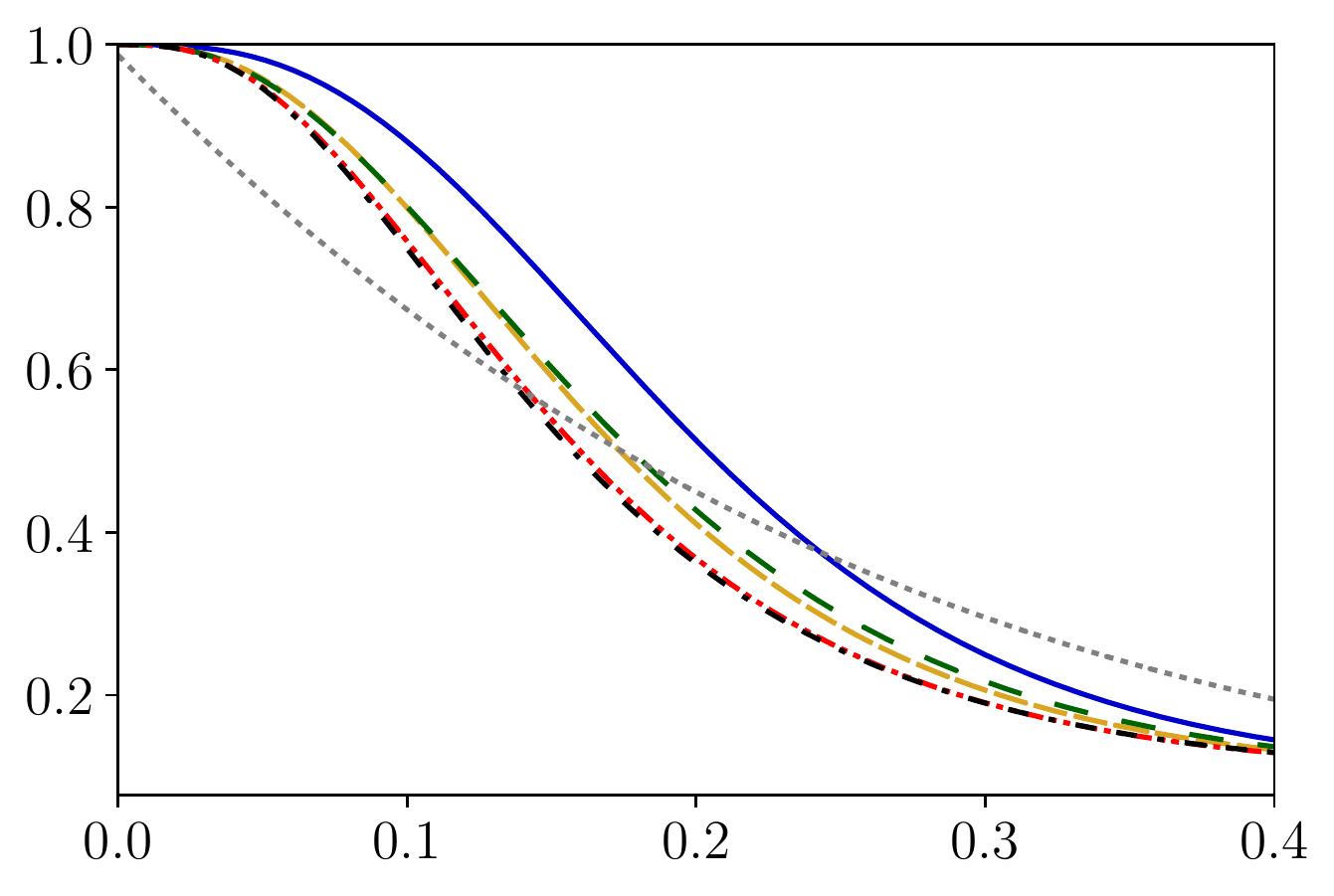}\\ 
\hspace{2.7em} 
$f $
\end{minipage}
\vspace{1em} 
\begin{minipage}{\columnwidth}
\begin{sideways} \hspace{6.4em} $P^\mathrm{distr}_{k_\mathrm{max}} $ \end{sideways}
\includegraphics[width= .9\columnwidth]{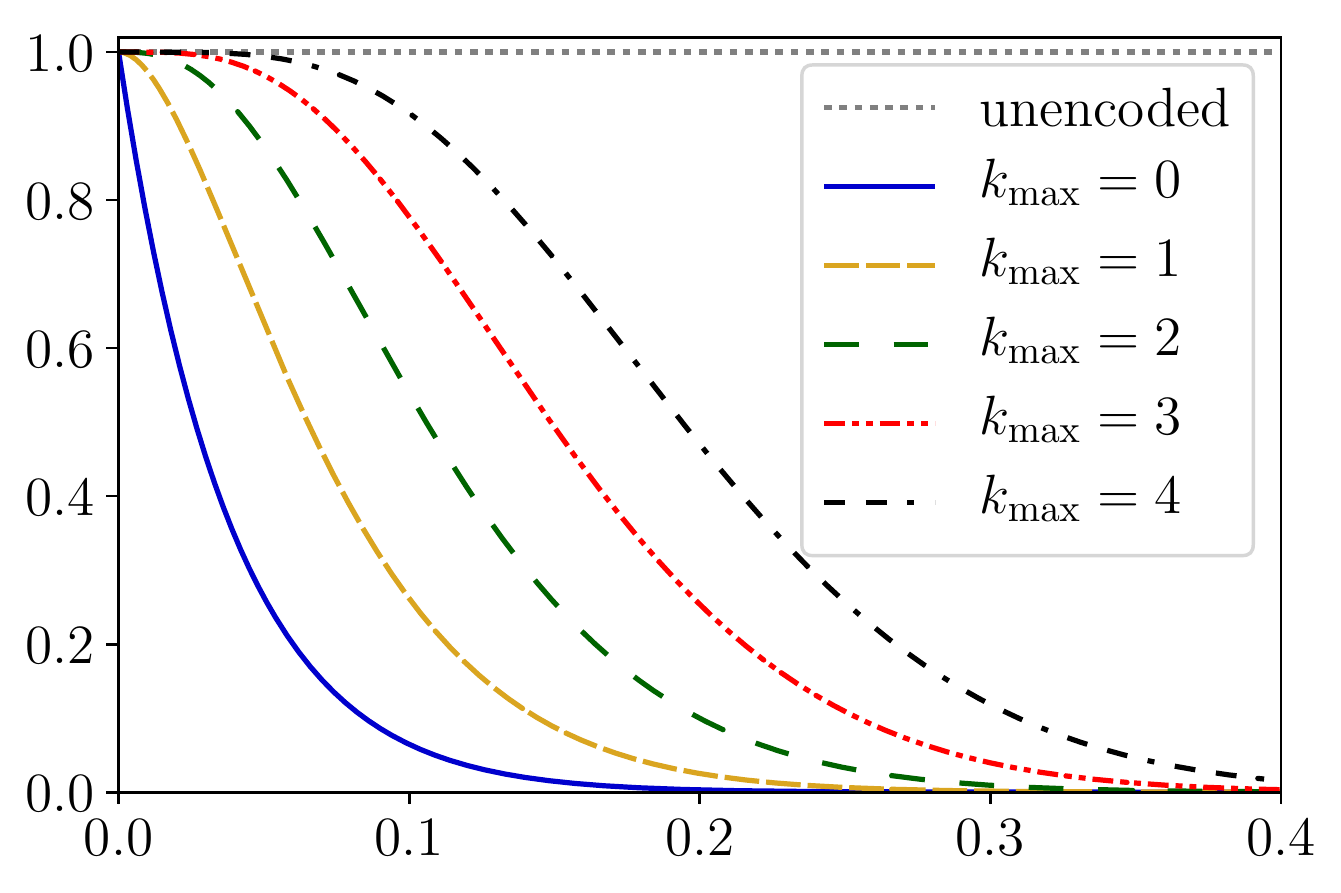}\\
\hspace{2.7em} $f$
\end{minipage} 
 \end{center}  
\caption{The fidelity, $F(k_\mathrm{max})$, of the distributed state (top), and the probability of successfully distributing the state~(bottom) as a function of unnoticed $(f_\mathrm{T})$ and noticed $(f_\mathrm{abs})$ transmission error rates $f:=f_\mathrm{T}= f_\mathrm{abs}$ for different abortion strategies. 
The repeater line has  $N=2$ stations, and measurement, gate, and storage  error rates are set to $f_\mathrm{M}=0.01$, $f_\mathrm{G}=0.001$,  and $f_\mathrm{S}=0.0001$, respectively.
The gray (dotted) curve shows the performance of an unencoded qudit with $D=13$.
The other curves show the performance of the $ \llbracket 13,1,7\rrbracket_{13}$  quantum polynomial code for different abortion conditions  $k_\mathrm{max}$.
If at any of the repeater stations the number of qudits marked as ``?'' is greater than $k_\mathrm{max}$, the protocol is aborted. }
\label{fig:abortion_thresholds} 
\end{figure}

As an example, we present this scheme for a $\llbracket13,1,7\rrbracket_{13}$ QECC.
The top plot in Fig.~\ref{fig:abortion_thresholds}
shows the behavior of the fidelity $F(k_\mathrm{max}) :=\sqrt{ \bra{\Psi}\rho(k_\mathrm{max} )\ket{\Psi}}$
of the distributed state $\rho(k_\mathrm{max})$ in terms of unnoticed and noticed transmission errors for various choices of $k_\mathrm{max}$.  
The bottom plot of \fig~\ref{fig:abortion_thresholds} shows the corresponding probability $P^\mathrm{distr}_{k_\mathrm{max}}$ of the protocol not being aborted, cf. \eq~\eqref{eq:P_distr} in Appendix~\ref{app:RL_QEC_abortion}.  
Due to a brute force approach, we can only solve  
$P^\mathrm{distr}_{k_\mathrm{max}}$ for a repeater line with $N=2$ repeater stations (including Bob), see Appendix~\ref{app:RL_QEC_abortion} for more details. 

In the following, we set $f:=f_\mathrm{T}=f_\mathrm{abs}$. 
First, consider the top plot of Fig.~\ref{fig:abortion_thresholds}. 
At $f=0$, the fidelity of the distributed state is $F(k_\mathrm{max})={1-10^{-5}}$, which is almost optimal. (For comparison, an unencoded repeater line yields a fidelity of $0.987$.)
Note that this is independent of $k_\mathrm{max}$ since no photons are lost.
The fidelities decrease in $f$ because of additional transmission errors. 
For $f>0$, they are arranged as
\begin{align} \label{eq:quality}
F(0)>F(1)\approx F(2)>F(3) \approx F(4).
\end{align} 
The difference between $F(0)$ and $F(1)$ is already significant for $f\approx 0.05$ because, in the case of an absorbed photon, a reduced $[ 12,1,6 ]_{13}$ code which can only correct up to 2 errors
%$\lfloor(6-1)/2\rfloor=2$, 
is used for $k_\mathrm{max}=1$, while for $k_\mathrm{max}=0$ the protocol is aborted if the original distance-7 code cannot be used.
Note that $F(1)$ and $F(2)$ are approximately the same because the $[11,1,5]_{13}$ code, which is additionally employed for $k_\mathrm{max}=2$, can correct as many errors as the $[12,1,6]_{13}$ code.
A similar argument holds for $F(3)\approx F(4)$. 
%Note that $F(1)$ and $F(2)$ intersect at $f\approx 0.089$.
%Likewise, $F(3)$ and $F(4)$ intersect at $f\approx0.32$.   
In the limit $f\rightarrow1$, all fidelities approach the worst-case value $F(k_\mathrm{max})=1/13\approx0.077$.

Now consider the bottom plot in Fig.~\ref{fig:abortion_thresholds}.
Note that the distribution probability does not depend on $f_\mathrm{T}$.  
At $f=0$, the probability of distributing a state is $P^\mathrm{distr}_{k_\mathrm{max}}=1$  because no qudits are lost and the protocol never aborts.
In total, $Nn=26$ photons are transmitted. 
For $k_\mathrm{max}=0$, the protocol is aborted if at least one photon is absorbed. 
This happens with probability $P^\mathrm{distr}_{0}=(1-f)^{26}$. 
This explains the rapid drop of the blue (solid) curve, $P^\mathrm{distr}_0$.
For higher $k_\mathrm{max}$, $P^\mathrm{distr}_{k_\mathrm{max}}$ decreases more slowly in $f$, cf. Eq.~\eqref{eq:P_distr},
as more photon losses are tolerated.\footnote{For example, for $k_\mathrm{max}=1$, the probability of distributing a state is
$P^\mathrm{distr}_{1}=(1-f)^{26}+ 26f(1-f)^{25}+13f^2(1-f)^{24}$, 
where ${26f(1-f)^{25}}$ accounts for the 26 events where exactly one photon is lost.
Similarly, the term $13f^2(1-f)^{24}$ accounts for 13 combinations of 2 lost photons which do not lead to an abortion.}
Thus, the distribution probabilities are ordered as 
\begin{align} \label{eq:quantity}
 P^\mathrm{distr}_{0}< P^\mathrm{distr}_{1}< P^\mathrm{distr}_{2}< P^\mathrm{distr}_{3}<  P^\mathrm{distr}_{4}.
\end{align}
 
Eqs. \eqref{eq:quantity} and \eqref{eq:quality} show the trade-off between the quantity and the quality of distributed states.
Naturally, one should not choose $k_\mathrm{max}$ to be odd (if $d$ is odd), since the fidelity is approximately that of $k_\mathrm{max}+1$ but the corresponding distribution probability is significantly lower.

\subsection{Optimizing the distributed entanglement}
\label{sec:RLQEC3}
  
Consider the following scenario:
Alice and Bob want to create an entangled state by using a qudit repeater line a single time.
The qudit can be encoded into an arbitrary $\llbracket 2d-1,1,d\rrbracket_D$ QECC with a fixed physical Hilbert space dimension $\dim (\mathcal H)=D^{2d-1}$.
(For example, if $\dim(\mathcal{H})=27$, Alice and Bob can choose between  $\llbracket 1,1,1 \rrbracket _{27}$ and $\llbracket 3,1,2\rrbracket _3$ encoding.)
They adjust the parameters $D$ and $d$ in order to maximize the logarithmic negativity
\begin{align}
\mathcal{E_N}(\rho) = \log_2( \vert\vert \rho^\mathrm{T_A} \vert \vert_{1} ),
\end{align}
where $\rho^\mathrm{T_A}$ is the partial transpose of $\rho$ with respect to Alice, and $\vert\vert\cdot\vert \vert _1$ is the trace norm~\cite{Negativity2002}.
The logarithmic negativity is an entanglement measure, and thus a quantifier of distributed resources.  

\begin{figure*} %[h!]
\begin{center}  
\begin{minipage}{.1\textwidth}  
\flushright \large
$D$ 
\end{minipage}
\begin{minipage}{.7\textwidth}   
\includegraphics[width= \textwidth]{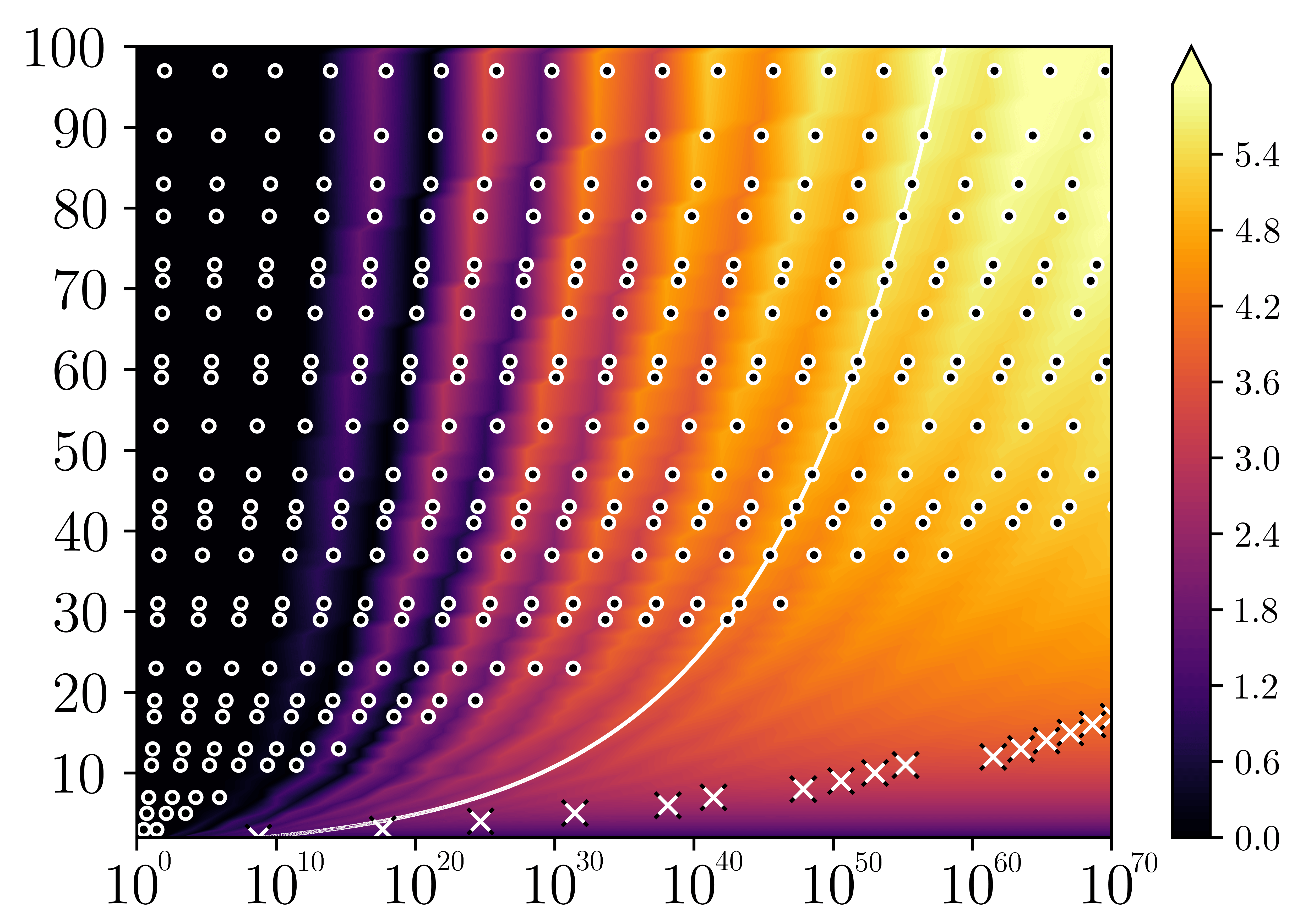}
\end{minipage} 
\begin{minipage}{.15\textwidth} 
\flushleft \large
$\mathcal{E_N}( \rho)$  
\end{minipage} \\ \large
\hspace{0em}$\mathrm{dim}(\mathcal{H})= D^{2d-1}$
\end{center}  
\caption {The logarithmic negativity $\mathcal{E_N}( \rho)$  of the state $\rho$ distributed via a repeater line with $N=50$ stations and encoded with a
$\llbracket 2d-1,1,d\rrbracket_ D$ code for varying qudit dimension~$D$ and code distance~$d$.  The ambient Hilbert space of a logical qudit is $\mathcal{H}$, i.e., if for example $D=100$ and $\dim\mathcal{H}=10^{70}$, one logical qudit is encoded into $35$ physical qudits. 
The transmission, measurement, gate, and storage  error rates are set to   $f_\mathrm{T}~=~0.05$, $f_\mathrm{M}=0.01$, $f_\mathrm{G}=0.001$, and $f_\mathrm{S}=0.0001$, respectively.
The dots represent parameters for which quantum polynomial codes exist, namely $1\le d\le (D+1)/2$ for every fixed prime $D$. The crosses represent the codes listed in Table~\ref{tab:max_logneg}. The white curve exemplary shows codes with constant code distance $d=15$.
}
\label{fig:logneg}   
\end{figure*}

In Fig.~\ref{fig:logneg}   
we show the logarithmic negativity $\mathcal{E_N}(\rho)$ for all $\llbracket 2d-1,1,d\rrbracket_D$ codes with  $2\le D \le 100$  and 
$D^{2d-1}\le 10^{70}$, the latter of which is the physical Hilbert space dimension of the system into which one logical qudit is encoded, i.e., $\mathcal{H}$ is the ambient Hilbert space of a logical qudit.
The state $\rho$ is distributed by a qudit repeater line with $N=50$ repeater stations (including Bob), where we assume that the error rates are independent of~$D$.\footnote{For storage, gate, and measurement errors, this assumption is probably not justified. 
However, experiments with time-bin qudits suggest that the transmission errors, which are the main error source, do not depend on $D$~\cite{Wong15}. 
For orbital angular momentum qudits, on the other hand, transmission errors increase with $D$~\cite{Leuchs18}. }   
We observe characteristic features in three different regions:
(i) For small code distances $d$, the repeater line distributes no entanglement, i.e., $\mathcal{E_N}(\rho)=0$. 
(ii) For small dimension $D$ and large code distances $d$, the logarithmic negativity is approximately optimal, i.e., $\mathcal{E_N}(\rho)\approx \log_2(D)$. This is the region below the crosses in Fig.~\ref{fig:logneg}. (iii) In between, $0\le\mathcal{E_N}(\rho)\le\log_2(D)$ holds. For a fixed dimension $D$, the  logarithmic negativity takes on values in an alternating fashion, governed by an overall trend to its maximum value, with increasing code distance~$d$.
{We will comment on} these three regions in the following: 

In region (i), for $d \in\{1,2,3,4,6\}$, too few errors can be corrected by the QECCs. 
Hence, the final state distributed to Alice and Bob lost any logarithmic negativity. Thus, it cannot even be used for entanglement distillation~\cite{Negativity2002}. The first non-zero logarithmic negativity arises for $d=5$, which shows that, for example, 
$\llbracket 9,1,5\rrbracket_D$ codes perform sufficiently well in the considered parameter region to distribute states that are entangled to some degree. 
For $d=6$, $\llbracket 10,1,6\rrbracket_D$ codes are used, which correct as many errors as the $\llbracket 9,1,5\rrbracket_D$ codes but rely on an additional physical qudit which also accumulates errors.
Overall, these perform worse, explaining the respective vanishing of the logarithmic negativity. 

In region (ii),  note that for a fixed dimension $D$ and sufficiently large distances $d\ge d_\mathrm{min}$, the distributed state $\rho$ is almost pure. Under these conditions, the logarithmic negativity is approximately that of a pure maximally entangled state $\ket{\Psi}\bra{\Psi}$, i.e., 
$\mathcal{E_N}(\rho)\approx\log_2(D)$. 
In Table~\ref{tab:max_logneg}, 
\begin{table}[h!] 
\begin{tabular}{|c||c|c|c|c|c|c|c|}\hline
$D$ 		& 2&3&4&5&6,7&$8,\ldots,11$&$12,\ldots,23$ \\ \hline  
$d_\mathrm{min}$&  15&19&21&23&25& 27& 29 \\ \hline
\end{tabular} 
\caption{Parameter of the smallest $\llbracket 2d_\mathrm{min}-1,1,d_\mathrm{min}\rrbracket_D$ codes for which in \fig\ref{fig:logneg} $\mathcal{E_N}(\rho)>0.99\times\log_2(D)$ holds. }
\label{tab:max_logneg}
\end{table} 
we show the values of $d_\mathrm{min}$ such that $\mathcal{E_N}(\rho)$ is above $0.99\times\log_2(D)$ for various dimensions $D$.   

Finally, in region (iii),  recall that codes with an odd code distance are beneficial for error correction. Precisely this argument explains 
the alternating values of $\mathcal{E_N}(\rho)$ for fixed dimension $D$. The overall trend to higher values of the logarithmic negativity is simply explained by the 
fact that the corresponding QECCs can correct more errors with increasing code distance. 

Overall, $\mathcal{E_N}(\rho)$ increases in the qudit dimension $D$ and the code distance $d$. Fixing either $d$ or $D$ and varying the other is not a fair comparison because the requirements to Alice and Bob also change, for example, the number of physical qudits $n=2d-1$ (for fixed $D$).
A better comparison is obtained if $\dim(\mathcal H)=D^n$ ($x$-axis in \fig~\ref{fig:logneg}) is fixed instead. This would be relevant if, for example a single ququad ($D=4, n=1$) is as expensive as two entangled qubits ($D=2, n=2$).  
Fig.~\ref{fig:logneg} shows that the optimal strategy depends on the chosen value of $\dim(\mathcal{H})$. 
If it is small, e.g., $10^{10}$, Alice and Bob should not increase $D$ too much.
If it is large, e.g., $10^{70}$, the logarithmic negativity is optimized for large $D$.
Even above the crosses, where distributed states are not  maximally entangled (for the corresponding $D$), $\mathcal{E_N}(\rho)$ still increases in $D$.
For experimental implementations, this is good because more quantum polynomial codes exist for larger $D$, while no QECCs are known in the region where  $\mathcal{E_N}(\rho)\approx \log_2(D)$.

\section{\label{sec:5}Conclusion and outlook}
The error probability tensor framework developed here is a useful tool for analyzing the propagation of generalized Pauli errors in quantum circuits composed of qudit Clifford gates.
%We have demonstrated this at the example of qudit repeater lines.
It enabled us to analytically derive the full error statistics of a state distributed via a qudit repeater line with arbitrary qudit dimension and  arbitrarily many repeater stations. 
Our analysis demonstrates the advantage of quantum repeaters with quantum error correction, as well as the trade-off between quality and preparation rate of distributed quantum states.
For a fixed number of provided degrees of freedom our analysis suggests that  higher-dimensional qudits can increase the amount of distributed entanglement.
In particular, we find that the amount of entanglement does increase in the qudit dimension only if sufficiently many errors can be corrected.
Fortunately, in the superior parameter region, explicit quantum error-correcting codes are  feasible in the form of quantum polynomial codes.
 
Experimentally, photonic qudits with physical qudit dimension up to the order of $10^5$ can be realized \cite{Wong15}. 
Missing key ingredients for the realization of the here discussed qudit repeaters are a procedure to encode logical states into a multi-photon system, as well as a way to physically implement the two-qudit controlled-phase gate between two physical photonic qudits. Gates between time-bin encoded qudits are especially desirable, as time-bin qudits are less prone to errors than, for example, orbital angular momentum qudits.

To conclude, we claim that the error probability tensor can be applied for analytical analyses of other quantum communication protocols, as they often only require Clifford gates.
 
\begin{acknowledgments}	
The authors thank Michael Epping and Liang Jiang for helpful discussions. 
We also thank Eric Sabo for feedback on the manuscript.
The circuit diagrams were typeset using the \LaTeX~package \texttt{Qcircuit.tex}~\cite{Qcircuit}.
%This project was not supported.
\end{acknowledgments}
 
% -----------------------------------------------------------------------------  

\appendix
\section{\label{app:QEC}Quantum polynomial codes}

An important class of higher-dimensional stabilizer QECCs are polynomial codes \cite{ GottesmanSecret, AharonovBenOr08, KKKS06, Cross08}. They have already proven to be useful in the context of qudit quantum repeaters \cite{Muralidharan17}. 
For every prime number $D$ and every number $d\le (D+1)/{2}$, there is a $\llbracket 2d-1,1,d\rrbracket _D$ quantum polynomial code. 
Here, we outline a specific subfamily of these QECCs.
  
Let $D\ge 3$ be an odd prime and let $d:=(D+1)/{2}$.
Consider the ${(d-1) \times D}$ parity check matrix  
\begin{align}    \label{def:parity_check}
    H &=  
    \begin{pmatrix}
    1&1 &1 &1 & \cdots & 1   \hspace*{1em} \\
    0&1 &2 &3 & \cdots & D-1 \hspace*{1.4em}  \\
    0&1 &2^2& 3^2 & \cdots &(D-1)^2 \hspace*{1em}  \\
\vdots&\vdots&\vdots&\vdots&\ddots&\vdots \hspace*{1em} \\
    0&1 & 2^{d-2} & 3^{d-2} & \cdots &(D-1)^{d-2}
    \end{pmatrix}  
\end{align}  
with entries $h_{j,k}:=k^j\in\FD$, where $0^0:=1$. 
%, and  $j\in\{0,\ldots,d-~1\}$, and $k\in\{0,\ldots,D-1\}$.
The vectors $\mathbf{h}_j:=~(k^j)_{0\le k \le D-1}~\in~(\FD)^D$ (rows of $H$) are mutually orthogonal, each of them is orthogonal to $\mathbf{i}:=(k^{d-1})_{0\le k \le D-1}$, and 
$\mathbf{i}\cdot \mathbf{i} =-1$  \cite{Handout}. 
Therefore, the % $2(d-1)=D-1$ 
operators $S_{j}^{X} := X^{\mathbf{h}_j}$,  $S_{j}^{Z} := Z^{\mathbf{h}_j}$ mutually commute, each of the $S^{X}_j$ and $S^Z_j$ commutes with  $X_\mathrm{L}:=
X^\mathbf{i} $  and $Z_\mathrm{L} := 
Z^\mathbf{-i}$, and $X_\mathrm{L}Z_\mathrm{L} =\omega ^{-1} Z_\mathrm{L}X_\mathrm{L}$  is fulfilled.
It follows that, $\mathcal{S}:= \left\langle S^{X}_j, S^Z_j \ \big\vert \ j \in \{0,\ldots, d-2 \} \right\rangle $ is an abelian subgroup of the qudit Pauli group on $D$ qudits. %(which does not contain any nontrivial operator proportional to the identity),
Therefore, $\mathcal{S}$ defines a QECC~\cite{Gottesman1999, Ghreghoriu_Standard_form} which encodes ${D-2(d-1) =1 }$ logical qudit with logical operators $X_\mathrm{L}$ and $Z_\mathrm{L}$.  
For each $ a\in \FD$, the logical code space has a  basis state
\begin{align} \label{def:RS_state}
\ket{a_\mathrm{L}} := \frac{1}{\sqrt{D^{d-1}}} \sum_{\substack{\lambda_0, \ldots, \lambda_{d-2} \in \FD \\  \lambda_{d-1}=a }}   
  \ket{f_{\boldsymbol{\lambda}}(0), 
  \ldots,f_{\boldsymbol{\lambda}}(D-1)},
\end{align}  
where for every vector $\boldsymbol{\lambda}:=(\lambda_0,\ldots,\lambda_{d-1}) \in (\FD)^d$ a corresponding polynomial is  defined as 
$f_{\boldsymbol{\lambda}}(T):= \lambda_0 + \lambda_1 T + 
\ldots + \lambda_{d-1}T^{d-1}$. 

The reason this constitutes a good QECC is the redundancy inherent in this construction: Since the polynomial $f_{\boldsymbol{\lambda}}$ is defined via its coefficients $\lambda_i$, one can reveal $f_{\boldsymbol{\lambda}}$ if $d$ evaluation values are known.
Let $k_0,\ldots,k_{d-1}\in \FD$ be  mutually distinct and define the $d\times d$ Vandermonde matrix $V:=(k_i^j)_{0\le i,j\le d-1}$ whose inverse is derived in \cite{VandermondeInverse}.
This reveals ${\boldsymbol{\lambda}} = V^{-1}(f_{\boldsymbol{\lambda}}(k_i))_{0\le i \le d-1 }$.
I.e., the system of linear equations 
\begin{align}\label{eq:LGS} 
\begin{rcases}  
f_{\boldsymbol{\lambda}}(0) &= \lambda_0 \\
f_{\boldsymbol{\lambda}}(1) &= \lambda_0+\lambda_1 + \ldots+ \lambda_{d-1} \\ 
&\hspace{.2em} \vdots \\
f_{\boldsymbol{\lambda}}(D-1) &= \lambda_0 
+\ldots+(D-1)^{d-1}\lambda_{d-1}  \hspace{1em}
\end{rcases} 
\end{align}
can be solved from only $d$ of its $D=2d-1$ equations.
Note that for these QECCs the C$Z$-gate is transversal in the sense that applying C$Z^{-1}$ to each pair of physical qudits within two logical qudits constitutes a logical C$Z$-gate~\cite{AharonovBenOr08}.

\section{\label{app:depol}Discretization of the depolarizing channel}
Here, we show that for each normalized $n$-qudit state~$\rho$ the relation
\begin{align} \label{app_dep_state}
    \frac{\mathbbm{1}}{D^n}= \frac{1}{D^{2n}}\sum_{\mathbf{r,s}\in (\ZDZ)^n} (X^\mathbf{r}Z^\mathbf{s})  \ \rho \ (X^\mathbf{r} Z^\mathbf{s})^\dagger
\end{align}
holds. 
This states that the depolarizing channel corresponds to some probability for discrete $X$- and $Z$-errors on the qudits -- an observation mentioned in \cite{EppingRouter}. 

To prove this, we expand the state $\rho$ in the computational basis,
\begin{align} 
  \rho &= \sum_{\mathbf{j,k}\in(\ZDZ)^n}  z_\mathbf{j,k} \ket{\mathbf{j}}\bra{\mathbf{k}},  
\end{align} 
and insert this expression and the expansion for Pauli operators, \eq~\eqref{eq:pauli_standard_form}, into the RHS of \eq~\eqref{app_dep_state}. 
By  orthonormality   we find
\begin{align} \nonumber
%\label{app_dep_a_line1}
 & \sum_{\mathbf{r,s}\in (\ZDZ)^n} (X^\mathbf{r}Z^\mathbf{s})  \ \rho \ (X^\mathbf{r} Z^\mathbf{s})^\dagger \\  
\label{app_dep_a_line4}
=& \sum_{\mathbf{r,s,j,k}\in (\ZDZ)^n} z_\mathbf{j,k} \ \omega^\mathbf{(j-k)\cdot s} \ket{\mathbf{j+r}} \bra{\mathbf{k+r}} .
\end{align}
  
Using the fact that complex roots sum up to zero, $\sum_{s\in (\ZDZ)^n} \omega^\mathbf{(l-m)s} = D^n \delta_\mathbf{l,m}  $ and
 $\mathrm{Tr}(\rho)=1$, the entries of the operator in \eq~\eqref{app_dep_a_line4} are given by
\begin{align} \nonumber
& \bra{\mathbf{l}}\left(\sum_{\mathbf{r,s,j,k}\in (\ZDZ)^n}  z_\mathbf{j,k}\ \omega^\mathbf{(j-k)\cdot s} \ket{\mathbf{j+r}} \bra{\mathbf{k+r}} \right) \ket{\mathbf{m}}\\ 
\nonumber
&= \sum_{\mathbf{r,s}\in( \ZDZ)^n}  z_\mathbf{l-r,m-r} \ \omega^\mathbf{(l-m)\cdot s} \\ 
\label{app_dep_b_line4} 
&= D^n \delta_\mathbf{l.m} \sum_{\mathbf{r}\in( \ZDZ)^n}  z_\mathbf{l-r,l-r}  
=  D^n \delta_\mathbf{l.m}   \ \ .
\end{align} 
Division by $D^{2n}$ yields \eq~\eqref{app_dep_state} and finishes the proof.

\section{\label{app:RL_ana}Error analysis of the qudit repeater line without QEC}

Here, we derive  \eq~\eqref{eq:ana_sol} from the main text for unencoded repeater lines. 
We begin with the error model in Sec.~\ref{app:RL_ana:model}.
In Sec. \ref{app:RL_ana:station}, we compute the error statics of the measurement at intermediate repeater stations, from which we derive the error statics of the distributed state in Sec. \ref{app:RL_ana:state}.

\subsection{Error model}
\label{app:RL_ana:model}
Since we do not assume a specific physical implementation of the qudits, 
all error sources are modeled by depolarizing channels. 
This is reasonable because for every error channel there is a worst case approximation by a depolarizing channel.
Nevertheless, our analysis can be adjusted to more specific error sources, if they can be modeled by  Pauli error channels with independent $X$- and $Z$-type errors.
Each faulty C$Z$-gate is modeled by a perfect gate followed by single-qudit error channels $\mathcal F_\mathrm{G}$ on each qudit. 
Every time a qudit is transmitted from one station to the next, it is acted upon by an error channel $\mathcal F_\mathrm{T}$. 
Each faulty measurement is modeled by an error channel $\mathcal F_\mathrm{M}$  
followed by a perfect measurement.
Moreover, Alice's qudit undergoes storage errors, modeled by $N$ channels  $\mathcal F_\mathrm{S}$.  
For simplicity, we assume that the respective error rates, $f_\mathrm{G}, f_\mathrm{T}, f_\mathrm{M}$, and $f_\mathrm{S}\in[0,1]$, are the same for each instance. 
 
Experimentally, we should also take preparation errors into account, but we find they are not a dominating source of error and do not include them here for the sake of simplicity. The framework presented in this paper can handle such errors if desired.

\subsection{Error statistics of measurements at intermediate repeater stations}
\label{app:RL_ana:station}

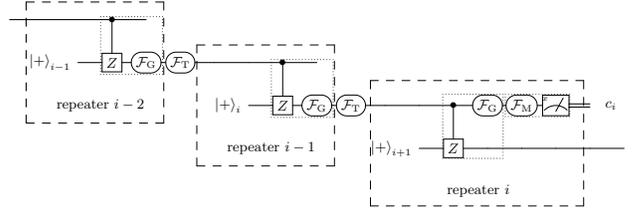
\begin{figure}[h!]
\begin{minipage}{\columnwidth} 
\centering 
\scalebox{0.6} 
{
\Qcircuit   @C=.2em @ R=.8em @! {
 & \qw& \qw& \ctrl{1}&\qw&  \\
 &&\lstick{\ket{+}_{i-1}}& \gate{Z} & \measure{ \mathcal F_\mathrm{G}} &\measure{ \mathcal F_\mathrm{T}} & \qw & \qw & \ctrl{1} & \qw & \\
 &&\text{\hspace{10mm}repeater $i-2$}&&&&&  \lstick{\ket{+}_i}& \gate{Z} & \measure{ \mathcal F_\mathrm{G}}& \measure{  \mathcal F_\mathrm{T}} & \qw & \qw & \ctrl{1} & \measure{\mathcal F_\mathrm{G}}& \measure{\mathcal F_\mathrm{M}}&\meterX  & \cw& \lstick{c_i}  \\
&&&&&  &&\text{\hspace{10mm}repeater $i-1$}&&&&&  \lstick{\ket{+}_{ i+1}}   & \gate{Z}&\qw&\qw&\qw& \qw&\qw\\
 &&&&&&&&&&&&&\text{\hspace{10mm} repeater $i$}&&&&& \ 
 \gategroup{1}{2}{3}{5}{2.4em}{--}  % repeater i-2
 \gategroup{2}{7}{4}{10}{2.4em}{--} % repeater i-1
 \gategroup{3}{12}{5}{17}{2em}{--}% repeater i
 \gategroup{1}{4}{2}{5}{0.2em}{.} %CZ i-2
 \gategroup{2}{9}{3}{10}{0.2em}{.} %CZ i-1
 \gategroup{3}{14}{4}{15}{0.1em}{.} %CZ i
 \gategroup{3}{16}{3}{17}{0.1em}{.} % meter i
  }  
 } 
\end{minipage}
\caption {Adapted from Fig.3 of \cite{EppingAnalysis}. Sources of undetected errors on the measurement outcome at repeater $i\in \{2,\ldots, N\}$.  If an $X$-error occurs on qudit $i-1$ (one gate, one transmission), it induces a $Z$-error on qudit $i$ across the C$Z$-gate. If this happens, or $Z$-errors directly occur on qudit $i$ (two gates, one transmission, one measurement), the measurement outcome might be erroneous.
}
\label{fig:RL_intermediate_errors} 
\end{figure}

As argued in \cite{EppingNetworks,EppingAnalysis}, errors propagate a distance of at most two repeater stations. 
Hence, an error on the measurement outcomes $c_i$, where $i\in \{2,\ldots,N\}$, can arise from six sources: $X$-errors from $\mathcal F_\mathrm{G}$ at repeater $i-2$ and from $\mathcal F_\mathrm{T}$ between repeaters $i-2$ and $i-1$ and $Z$-errors from $\mathcal F_G$ at repeater $i-1$, from $\mathcal F_\mathrm{T}$ between repeaters $i-1$ and $i$, as well as from $\mathcal F_\mathrm{G}$ and $\mathcal F_\mathrm{M}$ at repeater $i$, cf. \fig~\ref{fig:RL_intermediate_errors}.

Thus, the statistics of $Z$-errors right before the measurement are given by an error probability tensor $P' =(p'_s)$ with entries given by the tensor equation
\begin{align} \label{eq:intermediate_tensorequation}
  p'_s%^{(i)} 
&= M_s\,^a\ G_a\,^b\ T_b\,^c\ G_c\,^d\ T_d\,^e\ G_e\,^f\ p_f.
\end{align} 
Thereby, the error probability tensor is initialized to $P=(p_s)$ with $p_s=\delta_{s,0}$, and the tensors $M,G$, and $T$, which take measurement, gate, and transmission errors into account, respectively,  can be regarded as matrices of the form 
$\frac{f_\alpha}{D}O+ (1-f_\alpha)\mathbbm1$, 
where $ \alpha\in \{\mathrm{M,G,T}\}$ and
$O$ % = (1)_{1\le i,j\le D}$  
is the $D\times D$ matrix with all entries equal to~1. 
Because of $O^2\propto O$, the product of all matrices in \eq~\eqref{eq:intermediate_tensorequation}  
is also a matrix of the form $aO+b\mathbbm 1$, for some $a,b\in [0,1]$.
Expanding the product of matrices yields $b=(1-f_\mathrm{T})^2(1-f_\mathrm{G})^3(1-f_\mathrm{M})$, and the normalization $Da+b=1$ determines $a$. 
Hence, the entries of $P'$ are $p'_0=a+b$ and $p'_{s\neq 0}=a$,  
\begin{align} 
\nonumber
p'_0  &= 
\frac{1}{D} \left(1+(D-1)(1-f_\mathrm{T})^2(1-f_\mathrm{G})^3(1-f_\mathrm{M})   \right), \\
p'_{s\neq0} &=
\frac{1}{D} \left(1- (1-f_\mathrm{T})^2(1-f_\mathrm{G})^3(1-f_\mathrm{M})    \right)  .
\label{eq:intermediate_statistics} 
\end{align}

Note that, for measurements at station 1, the error probability tensor $p'_s= M_s\,^a\ G_a\,^b\ T_b\,^c\ G_c\,^d\  p_d$  differs slightly from \eq~\eqref{eq:intermediate_statistics} 
because, in contrast to qudits 1 to $N-1$, qudit~$\mathrm{A}$ was not exposed to the channels $\mathcal F_\mathrm{G}$ and $\mathcal F_\mathrm{T}$ before the first C$Z$-gate.
%as there are less error sources.

\subsection{Error statistics of the distributed state} 
\label{app:RL_ana:state}
Since the error statistics of measurements at different repeater stations   are independent of each other,\footnote{Error statistics of non-neighboring repeater stations are independent because errors propagate across at most one C$Z$-gate.
Moreover, for the depolarizing channel $f_{r,s}=f^X_rf^Z_s$ holds, where $f^X_r=\sum_sf_{r,s}$ and $f^Z_s=\sum_rf_{r,s}$, since $X$- and $Z$-errors are created independently.  Because of this and because only $Z$-errors lead to measurement errors, and only $X$-errors induce $Z$-errors across the C$Z$-gate, the error statistics of measurements at neighboring stations are  also independent.} 
the probability of a Pauli-frame recovery error follows from the joint probability of measurement errors at the respective repeater stations.
In particular, recovery errors give rise to an error channel on qudit $\mathrm{B}$ with coefficients $f_{r,s}=f^\mathrm{even}_rf^\mathrm{odd}_s$ determined by
\begin{align}\label{eq:f^even_tensorequation}
F^\mathrm{even}\,_a\, ^z &= F^{(2)}\,_a\, ^{b}\, F^{(4)}\, _b \,^c\, F^{(6)}\,_c\,^{d}  \ldots\, F^{(N)}\,_y \,^{  z}  , \\
F^\mathrm{odd}\,_a\, ^z &= F^{(1)}\,_a\, ^{b}\, F^{(3)}\, _b \,^c\, F^{(5)}\,_c\, ^{d}  \ldots\, F^{(N-1)}\,_y \,^{  z}  ,
\label{eq:f^odd_tensorequation}
\end{align} 
where $F^{(i)}\,_a\,^b:=f^{(i)}_{a-b}$ is the abbreviation introduced in Sec.~\ref{sec:3B2}.
Thereby,  $f^{(i)}_{a-b}$ comes from the the measurement error statistics of repeater station $i$, $f^{(i)}_s:=p'_{\pm s}$  as in \eq~\eqref{eq:intermediate_statistics},
and the signs of the indices come from \eq~\eqref{eq:postprocessed}.\footnote{The sign of the index of $p'_{\pm s}$ alternates in $i$ in the following way: $2:-$; $4:+$; $6:-$; $\ldots$ , and,  \ldots; $N-5:-$; $N-3:+$; $N-1:-$.
To be more explicit, e.g., $f_2 ^{(i)}= p'_{-2}$ and $f_4 ^{(i)}= p'_{+2}$.
Note that for depolarizing noise, $f^{(i)}_s=f^{(i)}_{-s}$. } 
The solution of \eq~\eqref{eq:f^even_tensorequation} and \eqref{eq:f^odd_tensorequation} is 
\begin{widetext}
 \begin{align} \nonumber 
 f^\mathrm{even}_0 &=   
 \frac{1}{D} \left(1+(D-1) \left( 
(1-f_\mathrm{G})^{{3N}/{2} }  (1-f_\mathrm{T})^{N} (1-f_\mathrm{M})^{{N}/{2}}  \right) \right),
\\ \label{eq:f_even}
 f^\mathrm{even}_{r\neq0} &= \frac{1}{D}\left(1- \left( 
(1-f_\mathrm{G})^{{3N}/{2}}  (1-f_\mathrm{T})^{N} (1-f_\mathrm{M})^{{N}/{2}}  \right)  \right)  ,
 \\ \nonumber 
  f^\mathrm{odd}_0 &= \frac{1}{D} \left(1+(D-1) \left( 
  (1-f_\mathrm{G})^{3N/2-1}  (1-f_\mathrm{T})^{N-1} (1-f_\mathrm{M})^{{N}/{2}}  \right) \right) ,
  \\
 f^\mathrm{odd}_{s\neq0} &=    
 \frac{1}{D}\left(1- \left( 
 (1-f_\mathrm{G})^{{3N}/{2}-1}  (1-f_\mathrm{T})^{N-1} (1-f_\mathrm{M}   )^{{N}/{2}}  \right)\right),
\end{align}
\end{widetext}
where $N\ge 2$ is even.
This noise, $f^\mathrm{even}_r$ and $f^\mathrm{odd}_s$, depolarizes along only the $X$- and $Z$-directions, respectively, as the other part of the (symmetrically) depolarizing noise vanishes since the $X$-errors commute with the $X$-measurements. 
Since the error statistics of the measurements are  independent of those of qudits $\mathrm{A}$ and $\mathrm{B}$, the error statistics of the distributed state are given by the truncated error probability tensor $\bar P=(p_{\mathrm{co}((0,r),(0,s))})$ with entries  given by
 \begin{align} \nonumber 
 & F^\mathrm{local}\,_r\,^a\,_s\,^b \,
 \underbrace{F^\mathrm{prop}\,_b\,^c\, F ^{\mathrm{odd}}\,_c\,^d}_{=: F^Z\,_b\,^d} \,
 \underbrace{F^{\mathrm{even}}\,_a\,^e }_{=: F^X\,_a\,^e} (\delta_{(d,e),(0,0)}) \\ 
 &= f^\mathrm{local}_{0,0}f^X_rf^Z_s + f^\mathrm{local}_\mathrm{err} \left( 1- f^X_rf^Z_s \right),
\label{eq:ana_sol2}
\end{align}
where  
\begin{align} \label{eq:sol_fZfX}
f^Z_k= f^X_k := f^\mathrm{even}_k 
\end{align}
with $f^\mathrm{even}_k$ as in \eq~\eqref{eq:f_even}.
This finishes the proof of \eq~\eqref{eq:ana_sol}  for unencoded repeaters.
Note that this solution can be obtained by taking normalization conditions into account, e.g., $1=f^Z_0+ (D-~1)f^Z_{s\neq0}$.
Also note that ${F}^\mathrm{local}\,_r\,^a\,_s\,^b $ is the error probability tensor of a depolarizing  channel on qudits $\mathrm{A}$ and $\mathrm{B}$ with strength parameter
\begin{align} \label{eq:sol_flocal}
f^\mathrm{local} := 1- (1-f_\mathrm{G})^2(1-f_\mathrm{S})^N,  
\end{align}
which defines $f^\mathrm{local}_{0,0}$ and $f^\mathrm{local}_\mathrm{err}:=f^\mathrm{local}_{(r,s)\neq(0,0)} $  via \eq~\eqref{eq:coefficients_dep}.  
Moreover, if an $X$-error occurs on qudit $N$, which can happen at the C$Z$-gate in repeater $N$ and during its transmission to Bob, this $X$- will induce a $Z$-error at qudit $\mathrm B$ across Bob's C$Z$-gate. 
This   is taken into account via $F^\mathrm{prop}\,_b\,^c$, the tensor corresponding to a $Z$-depolarizing error channel with strength parameter $f^\mathrm{prop}=1-(1-f_\mathrm{G})(1-f_\mathrm{T})$. 
Finally, note that \eq~\eqref{eq:ana_sol2} holds for all even $N\ge0$.

\section{\label{app:RL_QEC}Error analysis of the qudit repeater line with QEC}

Here, we derive  \eq~\eqref{eq:ana_sol} from the main text for encoded repeater lines. 
In particular, we generalize our error analysis to a qudit repeater line where each logical qudit consists of $n$ physical qudits.
For each physical qudit, we use the error model of Sec. \ref{app:RL_ana:model}. 
Consider an $\llbracket n,1,d\rrbracket_D$ QECC  with the following properties:
It allows transversal C$Z$-gates,  it can correct $X$- and $Z$-errors independently, 
and it has logical operators $X_\mathrm{L}=X^\mathbf{r}$ and $Z_\mathrm{L}=Z^\mathbf{s}$, where $\mathbf{r,s}\in(\ZDZ)^n$ have only invertible entries.
Note that, for example, quantum polynomial codes satisfy all of these properties.

\subsection{Error statistics of logical measurements at intermediate repeater stations}

At the logical $X$-measurement in a given repeater station, each of the $n$ physical qudits are individually measured in the $X$-basis.
Since the C$Z$-gate is transversal, errors do not spread across different blocks of physical qudits.
Due to this and our depolarizing noise model, 
the error statistics of individual physical qudits at the measurement in a repeater station are the same as in Sec.~\ref{app:RL_ana:station}.
The probability $p_{e_i}$ that an error $e_i\in \ZDZ$ occurs   
on one of the $n$ measurement outcomes is given in \eq~\eqref{eq:intermediate_statistics}. 
In particular, $p_1=\ldots=p_{(D-1)}$.
Thus, the probability of an error $\mathbf e =(e_1,\ldots,e_n)$ on the measurement outcomes at this station, is given by 
\begin{align}
p_\mathbf{e} &= \prod_{i=0}^{n} p_{e_i} = p_0^{n-H(\mathbf{e})}p_1 ^{H(\mathbf{e})},
\end{align}
where the Hamming weight $H(\mathbf{e})$ is the number of non-zero digits in $\mathbf{e}$.
Since an $\llbracket n,1,d\rrbracket_D$ code can correct up to  $\left\lfloor\frac{d-1}{2}\right\rfloor$ arbitrary single qudit errors, we can consider the following (not necessarily efficient) strategy:
If an error $\mathbf{e}$ with  $H(\mathbf{e})\le \left\lfloor\frac{d-1}{2}\right\rfloor$ occurs, we identify and correct it. 
If, on the other hand $H(\mathbf{e})> \left\lfloor\frac{d-1}{2}\right\rfloor$,  we assign a random digit to the logical measurement outcome. 
The probability that a correctable error  occurs is 
\begin{align} \label{eq:psucc_station}
p_\mathrm{cor} %:=&  \sum_{\substack{\mathbf{e} \in (\ZDZ)^n \\  H(\mathbf{e})  \le  \left\lfloor \frac{d-1}{2} \right\rfloor }}  p_\mathbf{e}  \\ \nonumber
: =&    \sum_{j=0 }^{ \left\lfloor \frac{d-1}{2} \right\rfloor }  
(D-1)^j {n \choose j}
\ p_0^{n-j}\ p_1^{j} , 
\end{align} 
where $(D-1)^j {n \choose j}$ is the number of vectors in $(\ZDZ)^n$ with exactly $j$ nonzero entries. %(since there are $(D-1)^j$ choices of $j$ nonzero entries, and $D$ choose $j$ positions to place them). 
It follows that the probability of a particular logical error $e_\mathrm{L}\neq 0$ is $p_\mathrm{guess}  = {1-p_\mathrm{cor}}/{D}$, which is independent of $e$.
The error correction is successful if either the error can be corrected or the occurred error was guessed. This happens with probability
\begin{align}
p_\mathrm{succ} = p_\mathrm{cor} + p_\mathrm{guess}  = \frac{1}{D}\left(1+(D-1)p_\mathrm{cor}\right).
\end{align} 
As before, all error rates are the same for each repeater station except for the first.

\subsection{Error statistics of the distributed logical state} 

We assume that Bob can perform a (perfect) round of stabilizer measurements before adjustming the Pauli-frame according to his and the repeater stations' measurement outcomes.
In this way, we  reduce the error statistics of the $2n$ physical to just $2$ logical qudits, while preserving all relevant information.

As per our error model, $X^r$- and $Z^s$-errors can be treated separately, and the respective probabilities are also independent of $r,s\neq0$. 
Due to this and the fact that the  logical state is stabilized by $X_\mathrm{A}\otimes Z_\mathrm{B}$ and $Z_\mathrm{A}\otimes X_\mathrm{B}$,  each $X$-error on one of Alice's physical qudits can be treated as  some $Z$-error on the corresponding physical qubit of Bob, and vice versa. (Here we need $X_\mathrm{L}=X^\mathbf{r}$ and $Z_\mathrm{L}=Z^\mathbf{s}$, where $\mathbf{r,s}\in(\ZDZ)^n$ have only invertible entries.)
 
As in the unencoded case, errors which are introduced by one gate and one transmission induce $Z$-errors on each physical qudit of Bob.
Additionally, local errors on the physical qudits are introduced by two gate and $N$ storage error sources.
The probability of an $X^r$- and $Z^s$-error on qudit $\mathrm{B}$ right before the stabilizer measurements is therefore,
$p^{X}\,_r= S^{(N)}\,_r\,^a\ G^{(2)}\,_a\,^b \ \delta_{b,0} $
and
$p^{Z}\,_s = S^{(N)}\,_s\,^a\ G^{(3)}\,_a\,^b\ T^{(1)}\,_b\,^c\ \delta_{c,0}$, 
respectively. Analogous to~\eq~\eqref{eq:intermediate_statistics} the solution of these tensor equations is
\begin{align} \nonumber
 p^X_0 &= \frac{1}{D}\left(1+(D-1)(1-f_\mathrm{G})^2(1-f_\mathrm{S})^N \right), \\ \nonumber
 p^X_{r\neq0} &= \frac{1}{D}\left(  1   -  (1-f_\mathrm{G})^2(1-f_\mathrm{S})^N \right) ,\\ \nonumber
 p^Z_0 &= \frac{1}{D}\left(1+(D-1)(1-f_\mathrm{G})^2(1-f_\mathrm{S})^N(1-f_\mathrm{T}) \right) ,\\ 
 p^Z_{s\neq0} &= \frac{1}{D}\left( 1  -  (1-f_\mathrm{G})^2(1-f_\mathrm{S})^N(1-f_\mathrm{T}) \right) .
\end{align} 
Employing the same correction strategy as before, the probability of a successful correction of $X$- and $Z$-errors is $p^X_\mathrm{succ}$ and $p^Z_\mathrm{succ}$, analogous to~\eq~\eqref{eq:psucc_station}, and the probability of a specific error is $p^{X,Z}_\mathrm{err}={(1- p^{X,Z}_\mathrm{succ}})/{D}$.
Hence, the probability of a logical $X^rZ^s$-error on Bob's qudit after the stabilizer measurement is
\begin{align}
 p%^\mathrm{cor}
 _{r,s} = \begin{cases}
 p_\mathrm{succ}^X p_\mathrm{succ}^Z & \text{ if } r=0,s=0 \\
 p_\mathrm{succ}^X p_\mathrm{err}^Z & \text{ if } r=0,s \neq0\\
 p_\mathrm{err}^X p_\mathrm{succ}^Z & \text{ if } r\neq0,s=0\\
 p_\mathrm{err}^X p_\mathrm{err}^Z & \text{ if } r\neq0,s\neq0 \hspace{1cm} .
                        \end{cases}
\end{align}
After the Pauli-frame adjustment, the error statistics of the distributed state finally become
\begin{align}\label{eq:ana_sol3}
 p_{\mathrm{co}((0,r),(0,s))} &= F^\mathrm{even}\,_r\,^a \ F^\mathrm{odd}\,_s\,^b \ p_{a,b},
\end{align}
where $\mathcal{F}^\mathrm{even}$ and $\mathcal{F}^\mathrm{odd}$ are the error channels defined in Eqs.~\eqref{eq:f^even_tensorequation} and \eqref{eq:f^odd_tensorequation}, respectively, but this time with the error rates of measurements on the \emph{logical} level. 
In this way, Eq.~\eqref{eq:ana_sol3} gets the same form as Eq.~\eqref{eq:ana_sol2}, the analytical result for the unencoded repeater line. 
In particular, this finishes the proof of \eq~\eqref{eq:ana_sol} for encoded repeaters.

\section{\label{app:RL_QEC_abortion}Error analysis of the qudit repeater line with QEC and abortion strategy} 
Here, we adapt our error analysis to repeaters with an abortion strategy, cf. Sec.~\ref{sec:RLQEC2}.
Recalling \eq~\eqref{def:f_abs}, a photon is absorbed during its transmission from repeater station $i-1 $ to $i$ with probability $f_\mathrm{abs}$,  and its absence is detected at the measurement of the corresponding qudit. 
The outcomes of such measurements %, which detected the loss, and of the corresponding measurement at the next
at repeater stations $i$ and $i+1$  are marked with a ``?''. 
The probability that  $k$ of the $n$ measurement outcomes are marked as ``?'' at the first repeater station is given by
\begin{align}
 P^\mathrm{first}_?(k) :={n \choose k} f_\mathrm{abs}^{k} (1-f_\mathrm{abs})^{n-k},
\end{align}
and, for every following repeater station by
\begin{align}
 P_?(k) := {n \choose k}(1-(1-f_\mathrm{abs})^2)^{k} ((1-f_\mathrm{abs})^2)^{n-k}.
\end{align} 
The outcomes which are marked as ``?'' are discarded, and the remaining $D$its constitute a classical error-correcting code with a Hamming distance of at least $d-k$ (equality if the original code does not inherit unnecessary redundancy). Hence, the logical measurement outcome is obtained by decoding the $n-k$ remaining physical outcomes according to a $[n-k, 1, d-k]_D$ error-correcting code. 
The probability of successfully correcting a given error with such a code is given by
\begin{align} \label{eq:psucc_station_abort}
p_\mathrm{cor,?}(k)   :=&    \sum_{j=0 }^{ \left\lfloor \frac{d-k-1}{2} \right\rfloor }  
(D-1)^j {n-k \choose j}
\ p_0^{n-k-j}\ p_1^{j}.
\end{align}   
The quality of the distributed state can be improved if the protocol is aborted if too many noticed errors occur.
Given that at most  $k_\mathrm{max} <d$ qudits are discarded, the probability that repeater station $i\in\{2,\ldots, N\}$ can correct an error is 
\begin{align} 
p_{\mathrm{cor}, k_\mathrm{max}}  &= \sum_{k=0}^{k_\mathrm{max}} \left( \frac{P_?(k)}{\sum_{k=0}^{k_\mathrm{max}} P_?(k)}\right) p_\mathrm{cor,?}(k) ,
\end{align}
and likewise for the first repeater station. 
This is because, in the case that the protocol is not aborted, which happens with probability $ \sum_{k=0}^{k_\mathrm{max}} P_?(k)$, the conditional probability that $k$ qudits are discarded is  $\frac{P_?(k)}{\sum_k  P_?(k)}$.

Assume that Alice and Bob do not know the number of noticed errors at the repeater stations. Then, the rest of the analysis is analogous to Appendix~\ref{app:RL_QEC}, where 
$p_\mathrm{cor}$ in \eq~\eqref{eq:psucc_station_abort} is replaced by $p_{\mathrm{cor},k_\mathrm{max}}$.
%The density matrix $\rho$ is then the 
%not known to Bob
%number of ? is not sent to Bob, only a two digits are sent to Bob, namely the latest version of $c_A$ and $c_B$.
%$=> <\rho> = \rho $

\subsection{The probability of distributing the state}
Here, we outline our approach for computing the probability of not aborting the protocol.
There are $N$ logical qudits transmitted, each of which is encoded into $n$ physical qudits (photons). 
Therefore, there are $Nn$ photons transmitted in total. 
The probability of successfully distributing the state (not aborting the protocol) is  
\begin{align}\label{eq:P_distr}
P^\text{distr}_{k_\mathrm{max}} :=  \sum_{m=0}^{Nn} \alpha(N,n,k_\mathrm{max};m) f_\mathrm{abs}^m(1-f_\mathrm{abs})^{Nn-m},
\end{align} 
where $\alpha(N,n,k_\mathrm{max};m)$ is the number of configurations with exactly $m$ absorbed photons, for which the protocol is not aborted (no logical measurement with more than  $k_\mathrm{max}$  physical outcomes marked as ``?'').
We formalize this combinatorial problem in the following way. 
To each possible configuration of absorbed photons, we assign an $N\times n$-matrix 
$A= (a_{i,j})$. 
If, at the  transmission from repeater $i-1$ to $i$, the $j^\mathrm{th}$ qudit is absorbed, the corresponding matrix entry is set to 0. Otherwise, it is set to 1.
With this, a matrix $A$ corresponds to a successful distribution attempt if,
for each of its rows $i\in\{1,\ldots,N\}$, the number of columns $j$ fulfilling $a_{i,j}a_{i-1,j}=0$ (the number of outcomes at repeater $i$ marked as ``?'') is at most $k_\mathrm{max}$, where we set $a_{0,j}=1$ for $1\le j\le n$ (as there is no transmission before repeater 1).
Thus, $\alpha(N,n,k_\mathrm{max};m)$ can be computed as the number of matrices, $A\in \mathbb F_2 ^{N\times n}$, which correspond to a successful distribution attempt with exactly $m$ zero entries.
For $N=2$ and $n=13$, we find the values of $\alpha(N,n,k_\mathrm{max};m) $ via a brute force computer search, see Table~\ref{tab:combinatorics}.

\begin{widetext}
\begin{center} 
\begin{table}[h!] 
\begin{tabular}{|c||c|c|c|c|c|c|c|c|c|c|}\hline
 & $m=0$ & $m=1$ & $m=2$& $m=3$& $m=4$& $m=5$& $m=6$& $m=7$& $m=8$ & $m=9$
\\ \hline \hline
	$k_\mathrm{max} = 0$&1&0&0&0&0&0&0&0&0&0
\\\hline$k_\mathrm{max} = 1$&1 & 26 & 13&0&0&0&0&0&0&0
\\\hline$k_\mathrm{max} = 2$&1 & 26 & 325 & 312 &    78 &0&0&0&0&0
\\\hline$k_\mathrm{max} = 3$&1 & 26 & 325 &2600 &  3510 &  1716 &   286& 0 & 0&0
\\\hline$k_\mathrm{max} = 4$&1 & 26 & 325 &2600 & 14950 & 24596 & 17446& 5720 &715&0
\\\hline
\end{tabular} 
\caption{The number of accepted configurations $\alpha(2,13,k_\mathrm{max};m)$. Using a brute force search over all matrices,  $A\in \mathbb F_2^{N\times n}$, we obtain the values   $\alpha(N,n,k_\mathrm{max};m)=\#\left\{ (a_{i,j}) 
\in \mathbb{F}_2^{N\times n}\ \big\vert \ m=\#\{a_{i,j} =0\} , \forall i \in\{1,\ldots, N\}
: k_\mathrm{max}\ge\#\{j\vert a_{i,j}a_{i-1,j}=0 \}  \right \}$, for $N=2$ and $n=13$.}
\label{tab:combinatorics}
\end{table} 
\end{center}
\end{widetext}

% ----------------------------------------------------------------------------- 
% 
%\bibliography{pra} 

\end{document}